\documentclass[12pt]{article}
\usepackage{epsfig}
\usepackage{amssymb}
        \newdimen\eqskip
        \newdimen\txtskip
        \baselineskip=\txtskip
        \newdimen\mysep                
        \newdimen\hmysep
\begin{document}
       
  \newcommand{\ccaption}[2]{
    \begin{center}
    \parbox{0.85\textwidth}{
      \caption[#1]{\small{{#2}}}
      }
    \end{center}
    }
\newcommand{\BS}{\bigskip}
\def    \be             {\begin{equation}}
\def    \ee             {\end{equation}}
\def    \beq             {\begin{equation}}
\def    \eeq             {\end{equation}}
\def    \ba             {\begin{eqnarray}}
\def    \ea             {\end{eqnarray}}
\def    \beqn           {\begin{eqnarray}}
\def    \eeqn           {\end{eqnarray}}
\def    \beeq           {\begin{eqnarray}}
\def    \eeeq           {\end{eqnarray}}
\def    \nn             {\nonumber}
\def    \=              {\;=\;}
\def    \frac           #1#2{{#1 \over #2}}
\def    \ret            {\\[\eqskip]}
\def    \ie             {{\em i.e.\/} }
\def    \eg             {{\em e.g.\/} }
\def    \bentarrow      {\:\raisebox{1.1ex}{\rlap{$\vert$}}\!\rightarrow}
\def    \rd             {{\mathrm d}}    
\def    \Im             {{\mathrm{Im}}}  
\def    \bra#1          {\mbox{$\langle #1 |$}}
\def    \ket#1          {\mbox{$| #1 \rangle$}}

\def    \kev            {\mbox{$\mathrm{keV}$}}
\def    \mev            {\mbox{$\mathrm{MeV}$}}
\def    \gev            {\mbox{$\mathrm{GeV}$}}


\def    \mq             {\mbox{$m_Q$}}  
\def    \mt             {\mbox{$m_t$}}  
\def    \mb             {\mbox{$m_b$}}  
\def    \mqq            {\mbox{$m_{Q\bar Q}$}}
\def    \mqqsq          {\mbox{$m^2_{Q\bar Q}$}}
\def    \pt             {\mbox{$p_T$}}
\def    \et             {\mbox{$E_T$}}
\def    \xt             {\mbox{$x_T$}}
\def    \xtsq           {\mbox{$x_T^2$}}
\def    \ptsq           {\mbox{$p^2_T$}}
\def    \etsq           {\mbox{$E^2_T$}}
        
\newcommand     \MSB            {\ifmmode {\overline{\rm MS}} \else 
                                 $\overline{\rm MS}$  \fi}
\def    \muf            {\mbox{$\mu_{\rm F}$}}
\def    \mug            {\mbox{$\mu_\gamma$}}
\def    \mufsq          {\mbox{$\mu^2_{\rm F}$}}
\def    \mur            {{\mbox{$\mu_{\rm R}$}}}
\def    \mursq          {\mbox{$\mu^2_{\rm R}$}}
\def    \mul            {{\mu_\Lambda}}
\def    \mulsq          {\mbox{$\mu^2_\Lambda$}}

\def    \bzero          {\mbox{$b_0$}}
\def    \as             {\ifmmode \alpha_s \else $\alpha_s$ \fi}
\def    \asb            {\mbox{$\alpha_s^{(b)}$}}
\def    \assq           {\mbox{$\alpha_s^2$}}
\def \oacube {\mbox{$ {\cal O}(\alpha_s^3)$}}
\def \oaemacube {\mbox{$ {\cal O}(\alpha\alpha_s^3)$}}
\def \oafour {\mbox{$ {\cal O} (\alpha_s^4)$}}
\def \oatwo {\mbox{$ {\cal O} (\alpha_s^2)$}}
\def \oaematwo {\mbox{$ {\cal O}(\alpha \alpha_s^2)$}}
\def \oaemas {\mbox{$ {\cal O}(\alpha \alpha_s)$}} 
\def \oas   {\mbox{$ {\cal O}(\alpha_s)$}}
\def\slash#1{{#1\!\!\!/}}
\def\rt1{\raisebox{-1ex}{\rlap{$\; \rho \to 1 \;\;$}}
\raisebox{.4ex}{$\;\; \;\;\simeq \;\;\;\;$}}
\def\ltap{\raisebox{-.5ex}{\rlap{$\,\sim\,$}} \raisebox{.5ex}{$\,<\,$}}
\def\gtap{\raisebox{-.5ex}{\rlap{$\,\sim\,$}} \raisebox{.5ex}{$\,>\,$}} 

\newcommand\LambdaQCD{\Lambda_{\scriptscriptstyle \rm QCD}}

\def\naive{na\"{\i}ve}
\def\asp{{\alpha_s}\over{\pi}}
\def\GE{\gamma_E}
\def\half{\frac{1}{2}}
\def\bom#1{\mbox{\bf{#1}}}
\begin{titlepage}
\nopagebreak
{\flushright{
        \begin{minipage}{5cm}
        CERN-TH/99-75\\
        Bicocca-FT-99-06\\
        DTP/99/34\\
        ITP-SB-99-7\\
        {\tt hep-ph/9903436}\\
        \end{minipage}        }
        
}
\vfill
\begin{center}
{\LARGE { \bf \sc 
Sudakov Resummation Effects in \\[0.3cm]
Prompt-Photon Hadroproduction\footnote{This work
was supported in part  by the EU Fourth Framework Programme ``Training and
Mobility  of Researchers'',  Network ``Quantum Chromodynamics and the Deep
Structure of Elementary Particles'', contract FMRX--CT98--0194 (DG 12 --
MIHT).}}

}
\vfill                                                       
{\bf      Stefano CATANI$^{(a)}$
     \footnote{On leave of absence from INFN, Firenze, Italy},
          Michelangelo L. MANGANO$^{(a)}$
     \footnote{On leave of absence from INFN, Pisa, Italy},
          Paolo NASON$^{(b)}$, \\
          Carlo OLEARI$^{(c)}$ and
          Werner VOGELSANG$^{(d)}$
}                                                                       
\\[0.5cm]
$^{(a)}$ CERN, Theoretical Physics Division, CH~1211 Geneva 23, Switzerland
\\
$^{(b)}$ INFN, Sezione di Milano, Italy
\\                                 
$^{(c)}$ Department of Physics, University of Durham, Durham, DH1 3LE, UK
\\                                 
$^{(d)}$ Institute for Theoretical Physics, State University of New York at
         Stony Brook, NY 11794-3840, USA

\end{center}                                   
\nopagebreak
\vfill
\begin{abstract} 
  We compute the effects of soft-gluon resummation, at the
  next-to-leading-logarithmic level, in the fixed-target
  hadroproduction cross section for prompt photons.  We find in
  general that the corrections to the fixed next-to-leading-order
  results are large for large renormalization scales, and small for
  small scales. This leads to a significant reduction of the
  scale dependence of the results for most experimental
  configurations of interest. We compare our results to the recent
  measurements by the E706 and UA6 collaborations.
\end{abstract}                                                
\vskip 1cm
CERN-TH/99-75\hfill \\
March 1999 \hfill  
\vfill       
\end{titlepage}
\section{Introduction}
\label{secintro}
The phenomenological interest of prompt-photon production in
fixed-target experiments~\cite{data,Huston98,UA6} resides mainly in
its use as a gluon probe in structure-function studies.  
Prompt-photon production
is historically our main source of information on the gluon 
parton density
at large~$x$ (e.g.
$x>0.2$)~\cite{Aurenche89,Vogelsang95,Martin98,CTEQ5,Huston95},
a region which has very
little influence on the evolution of the deep-inelastic-scattering
structure functions. This same region is relevant for hadron colliders
in production phenomena at very large transverse momenta, and thus its
understanding is crucial in order to disentangle possible new physics
signals from the QCD background. 

For example, a particularly
interesting problem has emerged in the past few years in the
production of large-transverse-energy (\et) jets at the Tevatron. An
excess over the QCD prediction has been reported by the CDF
collaboration~\cite{cdfjets}, for jets with $\et\gtrsim 350$~GeV. While
the excess has not been confirmed by the D$\emptyset$
data~\cite{d0jets}, it is of interest to study the uncertainty in the
high-\et\ tail of the jet distribution due to the gluon density
systematics, to see whether there is room for deviations as large as
those detected by CDF.  For example, a suitable modification of the
gluon density at large $x$ has been proposed
(CTEQ4HJ~\cite{Huston96}), which is consistent with the excess
observed by CDF. The study of the recent E706 prompt-photon
data~\cite{Huston98}, however, suggests that a consistent fit of the
large-$x_T$ ($x_T \equiv 2\et/\sqrt{S}$) rate is incompatible with the CTEQ4HJ
gluon density~\cite{Huston98}.  Moreover, both jet cross
sections and direct-photon cross sections at high transverse energy
are affected by soft-gluon effects. These effects should be understood
in both cases in order to be able to claim a discrepancy with QCD
predictions. In particular, these effects can be very important in the
direct-photon case, since the typical \et\ values probed are much
smaller than in the case of jet production at the Tevatron and,
therefore, the size of the running coupling $\as$ at the relevant
scales is bigger.

Comparisons between theory and prompt-photon experimental results have
been carried out recently in
Refs.~\cite{Vogelsang95,Martin98,Huston99,Aurenche98}. The recent E706
data~\cite{Huston98} seem to differ most from the next-to-leading order
calculation,
over the whole $x_T$ range.  In
Refs.~\cite{Huston98,Huston99,Martin98}, an attempt is made to fit the
E706 data by introducing an intrinsic transverse momentum of the
incoming partons with $\langle k_T^2\rangle \approx 1.2\sim 1.4\;{\rm
  GeV}^2$.  The precise details of how the intrinsic $k_T$ is
incorporated in the calculations, however, can significantly affect
the impact of these corrections, as shown by the large variations
reported in Ref.~\cite{Huston99}.  

The use of an
intrinsic-transverse-momentum model is sometimes motivated as a way of
estimating the effects of soft-gluon emission.  The most prominent
effect of soft-gluon emission in Drell-Yan pair production is the
generation of the characteristic transverse-momentum spectrum of the
lepton pair. This can be modeled with an appropriate intrinsic
transverse momentum of the incoming partons.  
As a matter of fact, the formalism
for soft-gluon resummation in Drell-Yan pair production can be shown
to merge, at very small transverse momenta, into some non-perturbative
intrinsic transverse momentum of the partons inside the
hadron~\cite{CSS}.  While this approach is not unreasonable when one
considers the transverse momentum of the produced {\em pair}, it can however
lead to inconsistencies for the problem 
of {\em single}-photon production.
In fact,
for example, it is quite clear that the photon $x_T$ spectrum at large
$x_T$ explores the kinematic region of $x\to 1$ in the parton
densities, which is certainly not the case for the transverse-momentum
distribution of a Drell-Yan pair. 
 Thus, as of now, a method for the inclusion of non-perturbative effects
in the resummed formulae for the high-$x_T$ limit of the inclusive photon
cross section is not available.
Furthermore, in the opposite limit of small $x_T$, it is the multiple emission
of hard (rather than soft) gluons that leads to a sizeable perturbative 
broadening of the
transverse momenta of the incoming partons~\cite{ktfact,ER}.

In this work, we consider the effect of soft-gluon resummation in
prompt-photon production near the threshold limit, that is to say
for $\xt \to 1$.  The theoretical evaluation of these effects, at the
next-to-leading logarithmic accuracy, has been carried out
independently in Refs.~\cite{Laenen98} and~\cite{CMN98}. We shall
review in the next section the necessary formalism, using the language
of Ref.~\cite{CMN98}. In the rest of the paper, we will present its
phenomenological applications, and we will thus discuss its numerical
implementation as well as its impact on physical cross sections.

As is well known, prompt-photon production takes place both by
hard-photon emission from initial- or final-state quarks (direct component), 
and 
by
collinear radiation from final-state partons.  This last mechanism is
not fully calculable in perturbation theory and, in fact, it
depends upon the photon fragmentation
function. Because of the large suppression of the fragmentation
function at large momentum fractions $z$, it is usually believed that this 
contribution
becomes irrelevant when $\xt$ increases.
Contrary to common wisdom,  
we shall instead show that
the very-large-$x_T$ behaviour of the direct and
of the fragmentation production processes is the same if the incoming
hadrons do not contain valence antiquarks, as in the case of $pN$
collisions.
Under these circumstances, resummation should therefore be performed
for the fragmentation emission too.  We will show, however, that in the
cases of practical interest the corrections 
due to the fragmentation processes are small, and we
shall limit our considerations to the hard-photon part.

The plan of the paper is as follows. In Section~\ref{secnotat} we
review the formalism for the resummation of threshold effects, and the
main formulas valid for the specific case of prompt-photon production.
There we also recall the main issues related to the inversion of the
resummed expressions from Mellin space back to the physical $x$ space.
In Section~\ref{Results} we study numerically the impact of the
resummation corrections. We explore the effects both at the parton
and hadron level, considering kinematical configurations and
distributions of phenomenological relevance for current experiments.
In particular, we concentrate on the study of the size of the
resummation corrections, and of the residual dependence on the choice
of renormalization and factorization scales. Section~\ref{datacomp}
contains a comparison between our results and the data from some
recent experiments. This does not want to be a comprehensive
phenomenological study, but a preliminary analysis of the impact of
our results on the comparison of theory and data.  Our conclusions,
and the outlook for future progress, are given in
Section~\ref{conclusions}. An Appendix collects some details of the
resummation formulas.

\section{Theoretical framework and notation}
\label{secnotat}                    
\subsection{Kinematics and cross section}
\label{seckinem}
We consider the inclusive production of a single
prompt photon in hadron collisions:
\beq
\label{procgamma}
H_1(P_1) + H_2(P_2) \to \gamma(p) + X \;\;.
\eeq
The colliding hadrons $H_1$ and
$H_2$ carry momenta $P_1^\nu$ and $P_2^\nu$, respectively. 
We parametrize the momenta in terms of light-cone coordinates:
\beq
P^\nu = ( P^+, {\bom P}_T , P^- ) \;, \;\;\;\; P^{\pm} \equiv \frac{1}{\sqrt 2}
\, (P^0 \pm P^3) \;.
\eeq
In their 
centre-of-mass frame, using massless kinematics, the momenta of the colliding
hadrons
have the following
light-cone coordinates
\beq
\label{mom}
P_1^\nu = {\sqrt \frac{S}{2}} \;(1,{\bom 0},0) \;, \;\;\;\;
P_2^\nu = {\sqrt \frac{S}{2}} \;(0,{\bom 0},1) \;\;,
\eeq
where $S= (P_1+P_2)^2$ is the centre-of-mass energy squared.
The photon momentum $p$ is thus parametrized as
\beq
\label{momp}
p^\nu = \left( \frac{E_T}{{\sqrt 2}} \,e^{y}, {\bom E}_T,
\frac{E_T}{{\sqrt 2}} \,e^{-y} \right) \;\;,
\eeq
where $E_T$ and $y$ are the transverse energy and the rapidity,
respectively. We also introduce
the customary scaling variable $x_T$ $(0 \leq x_T \leq 1)$:
\beq
\label{xt}
x_T = \frac{2 \,E_T}{\sqrt S} \;\;.
\eeq

In the present paper we are mostly
interested in the prompt-photon production cross section integrated
over $y$  at fixed $E_T$. According to perturbative QCD, the cross section
is given by the following factorization formula
\beeq
\label{1pxsgamma}
&&\frac{d\sigma_{\gamma}(x_T,E_T)}{d E_T} = \frac{1}{E_T^3} \sum_{a,b}
\int_0^1 dx_1 \;f_{a/H_1}(x_1,\mu_F^2)
\,\int_0^1 dx_2 \,\;f_{b/H_2}(x_2,\mu_F^2) \nonumber \\
&&\phantom{aaaaaaaa}\times \int_0^1 dx \left\{
\delta\!\left(x - \frac{x_T}{{\sqrt {x_1 x_2}}} \right) 
{\hat \sigma}_{ab\to {\gamma}}(x, \as(\mu^2); E_T^2, 
\mu^2, \mu_F^2, \mu_f^2) \right. \\
&&\phantom{aaa}+ \left. \sum_{c} \int_0^1 dz \;z^2 \;d_{c/\gamma}(z,\mu_f^2)
\;\delta\!\left(x - \frac{x_T}{z{\sqrt {x_1 x_2}}} \right)
\;{\hat \sigma}_{ab\to c}(x, \as(\mu^2); E_T^2, \mu^2, \mu_F^2, \mu_f^2)
\right\} \;\;. \nonumber
\eeeq
where $a,b,c$ denotes the parton indices $(a=q,{\bar q},g)$, and 
$f_{a/H_1}(x_1,\mu_F^2)$ and $f_{b/H_2}(x_1,\mu_F^2)$ are the
parton densities of the colliding hadrons, evaluated at the 
factorization scale $\mu_F$.
The first and the second term in the curly bracket on the right-hand side of 
Eq.~(\ref{1pxsgamma})
represent the {\em direct} and the {\em fragmentation} component
of the cross section, respectively. The fragmentation component involves the
parton fragmentation function $d_{c/\gamma}(z,\mu_f^2)$
of the observed photon at the factorization scale $\mu_f$, which, in general,
differs from the scale $\mu_F$ of the parton densities.

The {\em rescaled}\footnote{These functions are related to the partonic
differential cross sections by ${\hat \sigma}_{ab\to i} = E_T^3 \,
d{\hat \sigma}_{ab\to i} / dE_T$ ($i=\gamma,c$).}
partonic cross sections ${\hat \sigma}_{ab\to \gamma}$ 
and ${\hat \sigma}_{ab\to c}$ in Eq.~(\ref{1pxsgamma}) are
computable in QCD perturbation theory as power series expansions in the
running coupling $\as(\mu^2)$, $\mu$ being the renormalization scale
in the $\MSB$ renormalization scheme:
\beq
\label{pxsg}
{\hat \sigma}_{ab\to \gamma}(x, \as(\mu^2); E_T^2, \mu^2, \mu_F^2, \mu_f^2)
= \alpha \, \as(\mu^2) \left[
{\hat \sigma}_{ab\to d \gamma}^{(0)}(x) +
\sum_{n=1}^{\infty} \as^n(\mu^2) \, 
{\hat \sigma}_{ab\to \gamma}^{(n)}(x; E_T^2, \mu^2, \mu_F^2, \mu_f^2)
\right] \;,
\eeq
\beq
\label{pxsc}
{\hat \sigma}_{ab\to c}(x, \as(\mu^2); E_T^2, \mu^2, \mu_F^2, \mu_f^2)
= \as^2(\mu^2) \left[
{\hat \sigma}_{ab\to d c}^{(0)}(x) +
\sum_{n=1}^{\infty} \as^n(\mu^2) \, 
{\hat \sigma}_{ab\to c}^{(n)}(x; E_T^2, \mu^2, \mu_F^2, \mu_f^2)
\right] \;.
\eeq
Note that the ratio between the direct
and the fragmentation terms in Eqs.~(\ref{pxsg}) and (\ref{pxsc})
is of the order of $\alpha/\as$, where $\alpha$ is the fine structure constant.
This ratio is compensated by the
photon-fragmentation function $d_{c/\gamma}$, which (at least formally) is of 
the order
of $\alpha/\as$, so that direct and fragmentation components equally
contribute to Eq.~(\ref{1pxsgamma}). 

Throughout the paper we always use parton densities and parton 
fragmentation functions as defined in the $\MSB$ factorization scheme.
In general, we consider different values for the renormalization and 
factorization scales $\mu$, $\mu_F$, $\mu_f$, although we always assume that
all of 
them are of the order of the photon transverse energy $E_T$. 

The LO terms ${\hat \sigma}_{ab\to d \gamma}^{(0)}$ in Eq.~(\ref{pxsg})
are due to the following parton-scattering subprocesses at the
tree-level
\beq
\label{loproc}
q + {\bar q} \to g + \gamma \;, 
\quad
q + g \to q + \gamma \;,
\quad
{\bar q} + g \to {\bar q} + \gamma \;.
\eeq
Using our normalization, the two independent (non-vanishing) partonic cross 
sections for the direct component are:
\beq
\label{siqqgamma}
{\hat \sigma}_{q{\bar q} \to g\gamma}^{(0)}(x) = \pi \, 
e_q^2 \,\frac{C_F}{N_c} \;\frac{x^2}{\sqrt {1-x^2}}
\left(2 - x^2\right) 
\eeq
\beq 
\label{siqggamma}
{\hat \sigma}_{qg \to q\gamma}^{(0)}(x) = 
{\hat \sigma}_{{\bar q}g \to {\bar q}\gamma}^{(0)}(x) = \pi \,
e_q^2 \,\frac{1}{2N_c} \;\frac{x^2}{\sqrt {1-x^2}}
\left(1 + \frac{x^2}{4}\right) \;\;,
\eeq
where $e_q$ is the quark electric charge. Note that, having integrated
over the photon pseudorapidity, the 
expressions~(\ref{siqqgamma}) and~(\ref{siqggamma}) 
are even functions of the 
photon transverse energy $E_T$, i.e.\ they
depend on $x^2$ rather than on $x$. The NLO terms 
${\hat \sigma}_{ab\to \gamma}^{(1)}$ in Eq.~(\ref{pxsg}) were first computed in
Ref.~\cite{Aurenche}.

The partonic contributions ${\hat \sigma}_{ab\to c}$
to the fragmentation component of the cross section are exactly equal to those
of the single-hadron inclusive distribution. Note that, unlike in the case of
the direct component, {\em all} the parton-parton scattering subprocesses 
$ab\to c$
(i.e.\ including $ab=qq, gg$) contribute to the fragmentation component 
already at LO. The explicit calculation of ${\hat \sigma}_{ab\to c}$
up to NLO was performed in Ref.~\cite{ACGG}.

\begin{figure}
\begin{center}
\centerline{
\epsfig{file=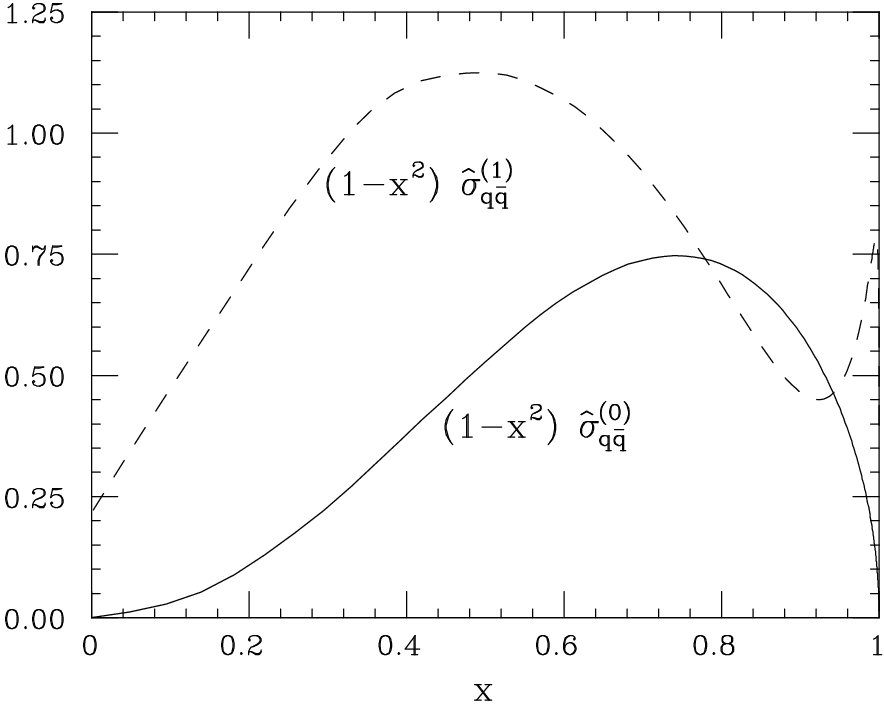,width=0.50\textwidth,clip=}\hfil
\epsfig{file=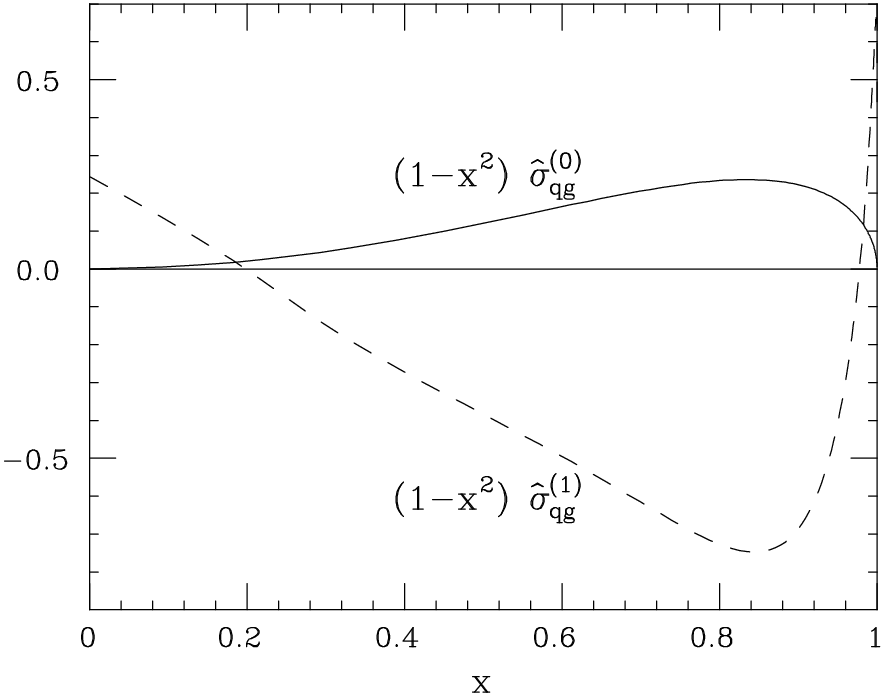,width=0.50\textwidth,clip=}                }
\ccaption{}{\label{fig:qalonlo} 
Behaviour of the LO and NLO terms ${\hat \sigma}_{ab\to d\gamma}^{(0)}$
and ${\hat \sigma}_{ab\to \gamma}^{(1)}$ (see Eq.~(\ref{pxsg}))
of the direct component of the prompt-photon cross-section.
The contributions of the partonic channels $ab=q{\bar q}$ (left) and 
$ab=qg$ (right) are rescaled by the factor $(1-x^2)$ and plotted as a function 
of $x$. The renormalization, factorization and fragmentation scales are all set
equal to $\mu^2=2\etsq$, and $e_q=1$.}
\end{center}
\end{figure}

The behaviour of the LO and NLO perturbative contributions to the 
direct component of the prompt-photon cross section
is shown in Fig.~\ref{fig:qalonlo}. 

The LO terms 
${\hat \sigma}_{q{\bar q}\to g \gamma}^{(0)}(x)$ and 
${\hat \sigma}_{qg\to q \gamma}^{(0)}(x)$ are both singular when 
$x \to 1$:
\beq
\label{lobeh}
{\hat \sigma}_{ab\to d \gamma}^{(0)}(x) \sim \frac{1}{\sqrt {1-x^2}} \;\;,
\;\;\;\; (x \to 1) \;\;,
\eeq
and they both vanish in the high-energy limit $x \to 0$. The
integrable singularity in Eq.~(\ref{lobeh}) is a typical phase-space effect,
while the vanishing behaviour at small-$x$ is due to the dominance of 
fermion (i.e.\ spin $1/2$) exchange in the $t$-channel.
 
Two new dynamical features appear at NLO. Near the threshold region $x \to 1$,
the NLO contributions are double-logarithmically enhanced,
\beq
\label{nlobeh}
{\hat \sigma}^{(1)}(x) \sim 
{\hat \sigma}^{(0)}(x) \; \ln^{2} (1-x) \;\;,  \;\;\;\;(x \to 1) \;\;,
\eeq
because the radiation of soft and, possibly, collinear partons is strongly 
inhibited by the kinematics. In the high-energy limit $x \to 0$, the partonic
cross sections ${\hat \sigma}_{ab \to \gamma}^{(1)}(x)$ approach constant
values~\cite{ER}: 
this Regge plateau follows from the fact that at NLO single-gluon (i.e.\ spin
1) $t$-channel exchange affects all the partonic subprocesses.
The behaviour of the partonic contributions ${\hat \sigma}_{ab \to
\gamma}^{(1)}(x)$ in the 
remaining intermediate region of $x$ has no straightforward physical 
interpretation 
(e.g.\ ${\hat \sigma}_{qg \to \gamma}^{(1)}(x)$ even becomes negative)
because it strongly depends on the scale-dependent
corrections already subtracted in the definition of the parton densities and
parton fragmentation functions.

Higher-order perturbative QCD corrections in the small-$x_T$ regime can  
systematically be computed by using the $k_{\perp}$-factorization 
approach~\cite{ktfact}, which consistently takes into account the {\em
perturbative} 
broadening of the transverse momenta of the incoming partons.

We are interested in this work in the behaviour of the QCD corrections near 
the partonic-threshold region $x \to 1$, i.e.\ when the transverse energy
$E_T$  
of the photon approaches the partonic centre-of-mass energy $\sqrt {x_1x_2S}$.
In this region, the  singularities in Eqs.~(\ref{lobeh}, \ref{nlobeh})
are enhanced by double-logarithmic corrections due to soft-gluon radiation and 
the higher-order cross
section contributions in Eqs.~(\ref{pxsg}, \ref{pxsc}) behave as
\beq
\label{hobeh}
{\hat \sigma}^{(n)}(x) \sim 
{\hat \sigma}^{(0)}(x) \left[ a_{n,2n} \; \ln^{2n} (1-x) + 
a_{n,2n-1} \ln^{2n-1} (1-x) 
+ \dots \right] \;\;.
\eeq
Resummation of these soft-gluon effects to all orders in perturbation theory
can be important to improve the reliability of the QCD predictions.

\subsection{$N$-moment space}
\label{Nsec}

The resummation program of soft-gluon contributions has to be carried 
out~\cite{Sterman, CT, CT2} in the Mellin-transform space, or $N$-space. 
Working in $N$-space, we can disentangle the soft-gluon effects in the parton
densities from those in the partonic cross section and we can
straightforwardly implement and factorize the kinematic constraints of
energy and longitudinal-momentum conservation. 

The latter point is particularly
relevant for soft-gluon resummation in hadron collisions~\cite{cmnt96}. 
Indeed, all-order soft-momentum recoil cannot exactly be taken into account
by directly working in $x$-space and the ensuing kinematics
approximation leads to (same-sign) factorially growing coefficients. 
This implies~\cite{cmnt96}
that no resummed logarithmic hierarchy can consistently be defined in $x$-space
(the classes of leading logs $\ln^{2n} (1-x)$, next-to-leading logs 
$\ln^{2n-1} (1-x)$ and so forth in Eq.~(\ref{hobeh}) are not separately
summable, because they lead to divergent and not integrable contributions
at $x=1$). On the contrary, no kinematics approximation (in the soft limit)
is required in $N$-space and 
the corresponding logarithmic hierarchy of $\ln N$-contributions 
is systematically well defined.

To work in $N$-space, it is convenient to consider the Mellin transform 
$\sigma_{\gamma, \,N}(E_T)$ of the dimensionless hadronic distribution 
$E_T^3  \, d\sigma_\gamma(x_T,E_T)/d E_T$. The $N$-moments
with respect to $x_T^2$ and at 
fixed $E_T$ are thus defined as follows:
\beq
\label{shn}
\sigma_{\gamma, \,N}(E_T) \equiv \int_0^1 dx_T^2 \;(x_T^2)^{N-1} 
\;E_T^3 \frac{d\sigma_\gamma(x_T,E_T)}{d E_T} \;\;.
\eeq 
In $N$-moment space, Eq.~(\ref{1pxsgamma}) takes a simple factorized form
\beeq
\label{1pxsngamma}
\sigma_{\gamma, \,N}(E_T) &=& \sum_{a,b}
f_{a/H_1,\,N+1}(\mu_F^2) \;f_{b/H_2,\,N+1}(\mu_F^2)
\nonumber \\
&\times & \Bigg\{
{\hat \sigma}_{ab\to \gamma, \;N}(\as(\mu^2); E_T^2, \mu^2, \mu_F^2, \mu_f^2) 
\\
&+& \left. \sum_c \,
{\hat \sigma}_{ab\to c, \;N}(\as(\mu^2); E_T^2, \mu^2, \mu_F^2, \mu_f^2)
\;d_{c/\gamma, \,2N+3}(\mu_f^2) \right\} \;\;, \nonumber
\eeeq
where we have introduced the customary $N$-moments $f_{a/H,\,N}$ and 
$d_{a/\gamma, \,N}$ of the parton
densities and parton fragmentation functions:
\beeq 
\label{pdm}
f_{a/H,\,N}(\mu^2) &\equiv& \int_0^1 dx \;x^{N-1} \;f_{a/H}(x,\mu^2) \;\;,
\\
\label{pffm}
d_{a/\gamma, \,N}(\mu^2) &\equiv& \int_0^1 dz \;z^{N-1} 
\;d_{a/\gamma}(z,\mu^2) \;\;.
\eeeq
Note that the $N$-moments of the partonic cross sections in 
Eq.~(\ref{1pxsngamma}) are again defined with respect to $x_T^2$:
\beq
\label{Ndef}
{\hat \sigma}_{ab\to \gamma, \;N}(\as(\mu^2); E_{T}^2, \mu^2, \mu_F^2, \mu_f^2)
\equiv 
\int_0^1 dx^2 \;(x^2)^{N-1} \;
{\hat \sigma}_{ab\to \gamma}(x,\as(\mu^2); E_{T}^2, \mu^2, \mu_F^2, \mu_f^2)
\;\;.
\eeq
The explicit expressions of the $N$-moments 
${\hat \sigma}_{q{\bar q}\to g\gamma, \;N}^{(0)}$,
${\hat \sigma}_{qg\to q\gamma, \;N}^{(0)}$
of the LO contributions in Eqs.~(\ref{siqqgamma}) and~(\ref{siqggamma}) were
obtained 
in Ref.~\cite{CMN98} and are recalled in Appendix~\ref{appa}.

Note also the pattern of moment indices in the various factors of 
Eq.~(\ref{1pxsngamma}), that is, $f_{a/H,\,N+1}$ for the parton densities and
$d_{c/\gamma, \,2N+3}$ for the parton fragmentation functions. This
non-trivial pattern follows from the conservation of the longitudinal
and transverse momenta.

The threshold region $x_T \to 1$ corresponds to the limit $N \to \infty$
in $N$-moment space. In this limit, the soft-gluon 
corrections~(\ref{hobeh}) to the higher-order contributions of the partonic
cross sections 
become
\beq
\label{hobehn}
{\hat \sigma}^{(n)}_N \sim 
{\hat \sigma}^{(0)}_N \; \left[ \; c_{n,2n} \;\ln^{2n} N 
 + c_{n,2n-1} \;\ln^{2n-1} N + \dots \right] \;\;.
\eeq
The resummation of the soft-gluon logarithmic corrections to all orders in
perturbation theory has been considered in Refs.~\cite{CMN98}
and~\cite{Laenen98}. 
In the following section we recall the main results.  

\subsection{Soft-gluon resummation at high $E_T$} 
\label{secnll}

In Ref.~\cite{CMN98} soft-gluon resummation has been performed in detail
for the various partonic channels that contribute to the {\em direct}
component of the prompt-photon cross section
$\sigma_{\gamma, \,N}(E_T)$ in Eq.~(\ref{1pxsngamma}).

We discuss first the large-$N$ behaviour of the partonic
cross sections ${\hat \sigma}_{ab\to \gamma, \;N}$ for 
the partonic channels $ab = q{\bar q}, qg, {\bar q}g$ that start to 
contribute at LO. These cross sections can be written as
\beq
\label{sigreslo}
{\hat \sigma}_{ab\to \gamma, \;N} = 
{\hat \sigma}_{ab\to \gamma, \;N}^{({\rm res})} \left[ 1 + {\cal O} (\as/N)
\right] \;,
\;\;\;ab = q{\bar q},\, qg, \,{\bar q}g \;,
\eeq
where ${\cal O}(\as/N)$ denotes terms that contribute beyond LO and are 
furthermore suppressed by a relative factor ${\cal O}(1/N)$ at large $N$.
The logarithmically-enhanced
soft-gluon corrections are included in the resummed
expressions ${\hat \sigma}_{ab\to \gamma, \;N}^{({\rm res})}$ and can 
be factorized with respect to the corresponding LO cross sections
${\hat \sigma}_{ab\to d\gamma, \;N}^{(0)}$.
The {\em all-order} resummation formulae are
\beeq
{\hat \sigma}_{q{\bar q}\to \gamma, \;N}^{({\rm res})}(\as(\mu^2); 
E_T^2, \mu^2, \mu_F^2, \mu_f^2) &=& \alpha \;\as(\mu^2) 
\;{\hat \sigma}_{q{\bar q}\to g\gamma, \;N}^{(0)} 
\;C_{q{\bar q} \to \gamma}(\as(\mu^2),Q^2/\mu^2;Q^2/\mu_F^2) \nonumber \\
\label{gammaresqq}
&\times&
\Delta_{N+1}^{q{\bar q} \to g \gamma}(\as(\mu^2),Q^2/\mu^2;Q^2/\mu_F^2)
\;\;, \\
{}\nonumber \\
{\hat \sigma}_{qg\to \gamma, \;N}^{({\rm res})}(\as(\mu^2); 
E_T^2, \mu^2, \mu_F^2, \mu_f^2) &=& \alpha \;\as(\mu^2) 
\;{\hat \sigma}_{qg\to q\gamma, \;N}^{(0)} 
\;C_{qg \to \gamma}(\as(\mu^2),Q^2/\mu^2;Q^2/\mu_F^2) \nonumber \\
\label{gammaresqg}
&\times& 
\Delta_{N+1}^{qg \to q \gamma}(\as(\mu^2),Q^2/\mu^2;Q^2/\mu_F^2)
\;\;, \\
{}\nonumber \\
\label{gammaresqbarg}
{\hat \sigma}_{{\bar q}g\to \gamma, \;N}^{({\rm res})}(\as(\mu^2); 
E_T^2, \mu^2, \mu_F^2, \mu_f^2) &=& 
{\hat \sigma}_{qg\to \gamma, \;N}^{({\rm res})}(\as(\mu^2); 
E_T^2, \mu^2, \mu_F^2, \mu_f^2)
\;\;,
\eeeq 
where 
\beq
\label{etscale}
Q^2 = 2 E_T^2 \;\;.
\eeq

The functions $C_{ab\to \gamma}(\as)$ in 
Eqs.~(\ref{gammaresqq}, \ref{gammaresqg}) do not depend on $N$.
Thus, the $\ln N$-dependence of the resummed cross sections 
is entirely embodied by the radiative factors
$\Delta_N^{ab \to d \gamma}$. They depend on the flavour
of the QCD partons $a, b, d$ involved in the LO hard-scattering
subprocess $a+b \to d + \gamma$ and can be expressed in an exponential form:
\beeq &&
\Delta_N^{ab \to d \gamma}\!\left(\as(\mu^2),\frac{Q^2}{\mu^2};
\frac{Q^2}{\mu_F^2}\right) =
\exp \Big\{ \ln N \; g_{ab}^{(1)}(b_0\as(\mu^2)\ln N) \nonumber \\ &&
\phantom{aaaaaa}+
g_{ab}^{(2)}(b_0\as(\mu^2)\ln N,Q^2/\mu^2;Q^2/\mu_F^2 )
\label{deltanll}
+ {\cal O}(\as(\as \ln N)^k) \Big\} \;\;,
\eeeq
where $b_0$ is the first coefficient of the QCD $\beta$-function
\beq
\label{betafirst}
b_0 = \frac{11 C_A - 4 T_R N_f}{12\pi}\;.
\eeq
Note that the functions $g^{(1)}$, $g^{(2)}$ and so forth in the exponent 
do not depend separately on $\as$ and $\ln N$. They are functions of the 
expansion variable $\lambda = b_0\as \ln N$ and vanish when $\lambda = 0$.
This means that the exponentiation structure in Eq.~(\ref{deltanll}) 
is not trivial and, in particular, that all the double logarithmic (DL) terms
$\as^n c_{n,2n} \ln^{2n} N$ in Eq.~(\ref{hobehn}) are taken into account
by simply exponentiating the lowest-order contribution 
$\as c_{1,2} \ln^2 N$.
The exponentiation in Eq.~(\ref{deltanll}) defines an improved 
perturbative expansion in the threshold region.
The function $\ln N \,g^{(1)}$
resums all the {\em leading} logarithmic (LL) contributions 
$\as^n \ln^{n+1} N$ in the exponent, 
$g^{(2)}$ contains the {\em next-to-leading}
logarithmic (NLL) terms $\as^n \ln^n N$, and so forth.
Once the functions
$g^{(k)}$ have been computed, we have a systematic perturbative
treatment of the region of $N$ where $\as \ln N \ltap 1$, which is much
larger than the domain $\as \ln^2 N \ll 1$ where the fixed-order calculation
in $\as$ is reliable.
 
The LL and NLL functions $g^{(1)}$ and $g^{(2)}$ in Eq.~(\ref{deltanll})
have been explicitly computed in Ref.~\cite{CMN98}. The LL functions $g^{(1)}$
are different for the $q{\bar q}$ and $qg$ partonic channels
of Eqs.~(\ref{gammaresqq}) and~(\ref{gammaresqg}) but they can be expressed
in terms of parton colour factors and a single (parton-independent) function
$h^{(1)}$:
\beqn
g_{q{\bar q}}^{(1)}(\lambda) &=& (2C_F - C_A) \;h^{(1)}(\lambda) +
C_A \;h^{(1)}(\lambda/2) \;, \;\;\; \;\;
\nonumber \\\label{g1fun}
g_{qg}^{(1)}(\lambda) &=& C_A \;h^{(1)}(\lambda) +
C_F \;h^{(1)}(\lambda/2) \;, 
\eeqn
with
\beq
\label{htll}
h^{(1)}(\lambda) =
\frac{1}{2\pi b_0 \lambda}\Bigl[ 2\lambda + (1-2\lambda)
\ln(1-2\lambda) \Bigr] \; . 
\eeq
The explicit expressions of the NLL functions $g^{(2)}$ are recalled in 
Appendix~\ref{appa}.

Note that the LL functions $g^{(1)}$ do not depend on the factorization scale
$\mu_F$. This dependence starts to appear only in the NLL functions $g^{(2)}$.
Note also the mismatch between the moment index of the radiative factor and that
of ${\hat \sigma}_{ab\to d\gamma, \;N}^{(0)}$ in Eqs.~(\ref{gammaresqq},
\ref{gammaresqg}): the former depends on $N+1$, 
like the parton densities in Eq.~(\ref{1pxsngamma}). The explicit
$\mu_F$-dependence of $g^{(2)}$ exactly matches the scale dependence of the
parton densities at large values of $N$. Thus, when (and only when) NLL 
resummation is included, we can expect~\cite{CMN98, BCMN}
better stabilization of the calculation of the
cross section at large $x_T$ with respect to variations of the factorization
scale $\mu_F$ (see Sec.~\ref{HLResults}).

The functions $C_{ab\to \gamma}(\as)$ in 
Eqs.~(\ref{gammaresqq}, \ref{gammaresqg})  
contain all the terms that are constant in the large-$N$ limit. They
are produced by 
hard {\em virtual} contributions and by subdominant (non-logarithmic)
soft corrections to the LO hard-scattering subprocesses.
These functions are computable as power series expansions in $\as$
\beeq
C_{ab\to \gamma}(\as(\mu^2),Q^2/\mu^2;Q^2/\mu_F^2) &=& 
1 + \sum_{n=1}^{+\infty} \; 
\left( \frac{\as(\mu^2)}{\pi} \right)^n \;
C_{ab\to \gamma}^{(n)}(Q^2/\mu^2;Q^2/\mu_F^2) \nn \\
\label{cgamma}
&=& 1 + \;\frac{\as(\mu^2)}{\pi} \;
C_{ab\to \gamma}^{(1)}(Q^2/\mu^2;Q^2/\mu_F^2) + {\cal O}(\as^2)
\;\;.
\eeeq
At present, we know only the first-order constant coefficients 
$C_{q{\bar q}\to \gamma}^{(1)}$ and $C_{qg\to \gamma}^{(1)}$ in 
Eqs.~(\ref{cgamma}, \ref{gammaresqq}, \ref{gammaresqg}). These
coefficients can be extracted~\cite{CMN98} from the complete NLO analytic 
results of Refs.~\cite{Aurenche, contogouris, gordon93}. Their values are
recalled in Appendix~\ref{appa}.

The inclusion of the $N$-independent function $C_{ab\to \gamma}(\as)$
in the resummed formulae does not affect the shape of the cross section
near threshold, but improves the soft-gluon resummation by fixing the
overall (perturbative) normalization of the logarithmic radiative factor.

We can explicitly show~\cite{CMN98, BCMN, CTTW} 
the theoretical improvement that is obtained by combining
the NLL radiative factor with the first-order coefficient $C_{ab\to
\gamma}^{(1)}$. 
Expanding the resummation formulae~(\ref{gammaresqq}, \ref{gammaresqg}) in
towers of logarithmic contributions as in Eq.~(\ref{hobehn}), we have
\beeq &&
 {\hat \sigma}_N^{({\rm res})}(\as;E_T^2,\mu^2,\mu_F^2) = 
\alpha \, \as \, 
{\hat \sigma}_N^{(0)}
\Big\{ 1 + \sum_{n=1}^{\infty} \as^n \Big[ c_{n,2n} \;\ln^{2n} N 
 + c_{n,2n-1}(E_T^2/\mu_F^2)  \;\ln^{2n-1} N
 \nonumber \\&&
+  c_{n,2n-2}(E_T^2/\mu_F^2,E_T^2/\mu^2) \;\ln^{2n-2} N
+ {\cal O}(\ln^{2n-3} N) \Big] \Big\} \;, \label{sigtow}
\eeeq
where $\as= \as(\mu^2)$. The dominant and next-to-dominant
coefficients $c_{n,2n}$ and $c_{n,2n-1}$ are controlled
by evaluating the radiative factor to NLL accuracy. When the NLL radiative
factor is supplemented with the coefficient $C_{ab\to \gamma}^{(1)}$, we can
correctly 
control also the coefficients $c_{n,2n-2}$. 
In particular, we can predict~\cite{CMN98} the large-$N$ behaviour of
the next-to-next-to-leading order 
(NNLO) cross sections 
${\hat \sigma}_{ab\to \gamma}^{(2)}$ in Eq.~(\ref{pxsg}) up to 
${\cal O}(\ln N)$. 

Note also that the coefficients $c_{n,2n}$ are scale independent and the
coefficients $c_{n,2n-1}$ depend on the sole factorization scale 
$\mu_F$. In the tower expansion~(\ref{sigtow}), the first terms that
explicitly depend on the renormalization scale $\mu$ (and on $\mu_F$, as well)
are those controlled by $c_{n,2n-2}$. Their dependence on $\mu$ is
obtained by combining that of $C_{ab\to
\gamma}^{(1)}(2E_T^2/\mu_F^2,2E_T^2/\mu^2)$
with that of the radiative factor at NLL order. The inclusion of the 
first-order constant coefficient $C_{ab\to \gamma}^{(1)}$ thus theoretically
stabilizes 
the resummed partonic cross section at large $x_T$ with respect to variations
of 
the renormalization scale. This scale dependence is numerically studied in
Sec.~\ref{HLResults}.

So far we have only considered the near-threshold behaviour  
of the partonic cross sections 
${\hat \sigma}_{q{\bar q}\to \gamma, \;N}$,
${\hat \sigma}_{qg\to \gamma, \;N}$, 
${\hat \sigma}_{{\bar q}g\to \gamma, \;N}$
in Eq.~(\ref{sigreslo}). The behaviour of other partonic channels 
$ab \to \gamma$ that 
contribute to the direct component of the prompt-photon cross section was 
discussed in Ref.~\cite{CMN98}. It turns out that the partonic channel
$ab = gg$ enters the resummed cross section only at next-to-next-to-leading
logarithmic (NNLL) accuracy and that all the other channels are relatively
suppressed in the same way as the correction ${\cal O}(\as/N)$ on the
right-hand side of Eq.~(\ref{sigreslo}).
Since we are interested in explicitly performing
soft-gluon resummation up to NLL order, we can limit ourselves to
considering the resummed expressions in 
Eqs.~(\ref{sigreslo})--(\ref{gammaresqbarg}). 

Detailed numerical studies of the resummed cross sections are presented in
Sec.~\ref{Results}. However, from the analytical results reviewed in this 
section, we may already anticipate that soft-gluon resummation increases 
the perturbative QCD predictions in the large-$x_T$ region. This 
conclusion can be argued by a simplified treatment within the DL approximation.
To DL accuracy, the exponent of the radiative factors in Eq.~(\ref{deltanll})
has to be expanded to its first order in $\as$, and we obtain
\beeq
\label{dlresqg''}
\frac{{\hat \sigma}_{qg\to \gamma, 
\;N}^{({\rm res})}}{{\hat \sigma}_{qg\to q\gamma, \;N}^{(0)}}
&\simeq& 
\exp \left\{ \left[ 2C_F + 2C_A -C_F \right]
\frac{\as}{2\pi} \ln^2 N \right\} 
=
\exp \left\{ \left( C_F + 2C_A \right)
\frac{\as}{2\pi} \ln^2 N \right\} > 1 \;\;, \\
&~& \nonumber \\
\label{dlresqq''}
\frac{{\hat \sigma}_{q{\bar q}\to \gamma, 
\;N}^{({\rm res})}}{{\hat \sigma}_{q{\bar q}\to q\gamma, \;N}^{(0)}} 
&\simeq&
\exp \left\{ \left[ 2C_F+ 2C_F - C_A \right]
\frac{\as}{2\pi} \ln^2 N \right\} =
\exp \left\{ \left( 4C_F - C_A \right)
\frac{\as}{2\pi} \ln^2 N \right\} > 1
\;\;, \\
&~& \nonumber \\
\label{dlrat'}
\frac{{\hat \sigma}_{qg\to \gamma, \;N}^{({\rm res})}}
{{\hat \sigma}_{q{\bar q}\to \gamma, \;N}^{({\rm res})}} &\simeq&
\frac{{\hat \sigma}_{qg\to q\gamma, \;N}^{(0)}}
{{\hat \sigma}_{q{\bar q}\to q\gamma, \;N}^{(0)}} \;
\exp \left\{ 3(C_A - C_F) 
\frac{\as}{2\pi} \ln^2 N \right\} >
\frac{{\hat \sigma}_{qg\to q\gamma, \;N}^{(0)}}
{{\hat \sigma}_{q{\bar q}\to q\gamma, \;N}^{(0)}} \;\;.
\eeeq
For the sake of completeness, 
in the square bracket on the right-hand side of
Eqs.~(\ref{dlresqg''}) and~(\ref{dlresqq''})
we have explicitly separated
the positive contributions coming from the initial-state
partons and the negative contribution from the final-state recoil.
From these equations we see that the resummed partonic cross sections 
${\hat \sigma}_{q{\bar q}\to \gamma, \;N}^{({\rm res})}$
and ${\hat \sigma}_{qg\to \gamma, \;N}^{({\rm res})}$
are both enhanced with respect to their LO approximations 
${\hat \sigma}_{q{\bar q}\to g\gamma, \;N}^{(0)}$,
${\hat \sigma}_{qg\to q\gamma, \;N}^{(0)}$. Moreover, the enhancement in the
$qg$ partonic channel is larger than that in the $q{\bar q}$ channel.
We refer the reader to Ref.~\cite{CMN98} for a discussion on the
physical origin of this behaviour.

\subsection{Fragmentation component}
\label{fragcomp}
We can now comment on the large-$E_T$ behaviour of the fragmentation component
of the prompt-photon cross section, by comparing the direct and fragmentation 
contributions in Eq.~(\ref{1pxsngamma}).

The partonic cross sections
${\hat \sigma}_{ab\to \gamma, \;N}$ and ${\hat \sigma}_{ab\to c, \;N}$ 
have the same large-$N$ behaviour, but,
owing to the hard
(although collinear) emission always involved in any splitting process
$c \to \gamma + X$, the photon fragmentation function $d_{c/\gamma, \,N}$
is of the order of $1/N$. Therefore, in the curly bracket on the right-hand
side of Eq.~(\ref{1pxsngamma})
the fragmentation component is formally suppressed by
a factor of $1/N$ with respect to the direct component. 
This suppression is consistent with the fact that the 
resummed partonic cross sections for the direct processes (see the 
right-hand side of Eqs.~(\ref{gammaresqq}) and~(\ref{gammaresqg})) 
turn out to be independent of the photon fragmentation scale $\mu_f$.

This argument shows that, in many cases, the fragmentation contributions 
are subdominant near threshold and, thus, they can be neglected in resummed
calculations at large $x_T$.

The caveat `in many cases' in the above conclusion regards the fact that the
argument applies to the partonic contributions in the curly bracket
of Eq.~(\ref{1pxsngamma}). In other words, the argument assumes
that all the different initial-state partonic channels 
$ab$ give comparable contributions to the hadronic cross section.
This is not always true once the effect of the parton densities is included.
 
A relevant exception is indeed the case of prompt-photon production in 
proton-nucleon collisions. Owing to the low antiquark content of
the colliding hadrons, the hadronic cross section is 
mostly due to
the partonic channels
$ab = qg$ and $ab = qq$:
\beq
\label{xsngammapN}
\sigma_{\gamma, \,N}(E_T) \sim  \sigma_{\gamma, \,N}^{qg}(E_T)
+ \sigma_{\gamma, \,N}^{qq}(E_T) \;\;, \;\;\;(pN \;{\rm collisions}) \;.
\eeq
As for the $qg$ initial-state contribution $\sigma_{\gamma, \,N}^{qg}(E_T)$, 
we can use the above argument to conclude that its direct component dominates
at large $E_T$. Setting all the scales equal to $E_T$, for the sake of
simplicity, we can write:
\beq
\label{xsngammapNqg}
\sigma_{\gamma, \,N}^{qg}(E_T) \sim 
\sigma_{\gamma, \,N}^{qg \,({\rm dir})}(E_T) \sim
f_{q,\,N+1}(E_T^2) \;f_{g,\,N+1}(E_T^2) \;
{\hat \sigma}_{qg\to \gamma, \;N}(\as(E_T^2))
\;\;.
\eeq
However, in the case of the $qq$ initial state, the direct component enters
only at NLO and, thus, the cross section is dominated by the fragmentation part
and, in particular, by photon fragmentation from a final-state quark
of the LO scattering subprocess $q+q \to q+q$. We can write:
\beq
\label{xsngammapNqq}
\sigma_{\gamma, \,N}^{qq}(E_T)  \sim
\sigma_{\gamma, \,N}^{qq \,({\rm frag})}(E_T)  \sim
f_{q,\,N+1}(E_T^2) \;f_{q,\,N+1}(E_T^2)
\;{\hat \sigma}_{qq\to q, \;N}(\as(E_T^2))
\;d_{q/\gamma, \,2N+3}(E_T^2)
\;\;.
\eeq

Taking the ratio of the two initial-state contributions and replacing 
${\hat \sigma}_{qq\to q, \;N}(\as)$ and
${\hat \sigma}_{qg\to \gamma, \;N}(\as)$ by their LO contributions 
${\hat \sigma}^{(0)}$ in Eqs.~(\ref{pxsg}, \ref{pxsc}), we obtain
\beq
\label{qgqqratio}
\frac{\sigma_{\gamma, \,N}^{qq \,({\rm frag})}(E_T)}{\sigma_{\gamma, 
\,N}^{qg\,({\rm dir})}(E_T)} \sim 
\frac{f_{q,\,N+1}(E_T^2)}{f_{g,\,N+1}(E_T^2)} \;d_{q/\gamma, \,2N+3}(E_T^2)
\; \frac{\as(E_T^2)}{\alpha} 
\;\frac{{\hat \sigma}_{qq\to qq, \;N}^{(0)}}{{\hat \sigma}_{qg\to q\gamma, 
\;N}^{(0)}} \;\;.
\eeq
The factor $\as/\alpha$ on the right-hand side is compensated by the behaviour
of the photon fragmentation function $d_{q/\gamma, \,2N+3} \propto \alpha/\as$.
In the large-$N$ limit, the ratio of the LO partonic contributions 
${\hat \sigma}^{(0)}$ is constant and, thus, the fragmentation function
produces an ${\cal O}(1/N)$-suppression factor. Nonetheless, this suppression
can be balanced by the parton density contribution 
$f_{q,\,N+1}/f_{g,\,N+1}$ since, at large $x$, the gluon density is typically
softer than the quark density.
As a matter of fact, using the Altarelli-Parisi evolution equation at LO and
under reasonable assumptions on the large-$x$ behaviour
of the parton densities at the initial evolution scale, 
it is easy to show that we have the following asymptotic behaviour
at very large values of $N$ and of the evolution scale:
\ba
  \label{eq:largeNfg}
f_{g,\,N} \sim \frac{1}{N\, \ln N} \; f_{q,\,N}  \;\;, \\
  \label{eq:largeNdg}
d_{q/\gamma, \,N} \sim \frac{1}{N\, \ln N} \frac{\alpha}{\as} \;\;.
\ea
Combining these results with Eq.~(\ref{qgqqratio}), in
the large-$N$ limit we therefore get
\be
\label{eq:fraglargeN}
\frac{\sigma_{\gamma, \,N}^{qq \,({\rm frag})}(E_T)}{\sigma_{\gamma, 
\,N}^{qg \,({\rm dir})}(E_T)} \stackrel{N\to\infty}{\to} \; {\rm
constant} \;\;.
\ee

This discussion shows that, in the case of large-$x_T$ prompt-photon production
in $pN$ collisions, the contribution
of the fragmentation component to the hadronic cross section can become
comparable to that of the direct component. 

Although on the right-hand side of Eq.~(\ref{qgqqratio}) we have approximated
the partonic contribution 
${\hat \sigma}_{qq\to q, \;N}(\as)/{\hat \sigma}_{qg\to \gamma, \;N}(\as)$
by its LO expansion, the inclusion of higher-order terms and, in particular,
of resummation effects does not substantially modify the conclusion.

Performing soft-gluon resummation in the partonic cross sections
${\hat \sigma}_{ab\to c, \;N}$
of the fragmentation component, we can write an expression that is analogous to
Eq.~(\ref{sigreslo}): 
\beq
\label{sigreslof}
{\hat \sigma}_{ab\to c, \;N} = 
{\hat \sigma}_{ab\to c, \;N}^{({\rm res})} \left[ 1 + {\cal O} (\as/N)
\right] \;.
\eeq 
Limiting our treatment to the LL accuracy, the resummed cross section is given 
by~\cite{CMN98,Laenen98}
\beq
\label{gammaresf}
{\hat \sigma}_{ab\to c, \;N}^{({\rm res})}(\as(\mu^2); 
E_T^2, \mu^2, \mu_F^2, \mu_f^2) \simeq \as^2(\mu^2) 
\;{\hat \sigma}_{ab\to dc, \;N}^{(0)} \;
\Delta_{N+1}^{ab \to dc}(\as(\mu^2),Q^2/\mu^2;Q^2/\mu_F^2,Q^2/\mu_f^2) \;\;. 
\eeq
The radiative factor is 
\beq 
\label{deltanllf}
\Delta_N^{ab \to d c}\!\left(\as(\mu^2),\frac{Q^2}{\mu^2};
\frac{Q^2}{\mu_F^2},\frac{Q^2}{\mu_f^2} \right) =
\exp \Big\{ \ln N \; g_{ab\to dc}^{(1)}(b_0\as(\mu^2)\ln N) 
+ {\cal O}((\as \ln N)^k) \Big\} \;\;,
\eeq
where the LL function $g_{ab\to dc}^{(1)}$ is analogous to those in 
Eq.~(\ref{g1fun}) and it can be expressed in terms of the colour charges
$C_a$ ($C_a=C_F$, if the parton $a$ is a quark, and $C_a=C_A$, if $a$ is
a gluon) of the partons involved in the LO hard-scattering subprocess:
\beq
g_{ab\to dc}^{(1)}(\lambda) = (C_a + C_b + C_c - C_d) \;h^{(1)}(\lambda) +
C_d \;h^{(1)}(\lambda/2) \;. 
\eeq

In particular, for the $q + q\to q + q$ channel we have
\beq
\label{g1qqqq}
g_{qq\to qq}^{(1)}(\lambda) = 2 C_F \;h^{(1)}(\lambda) +
C_F \;h^{(1)}(\lambda/2) \;, 
\eeq
which is very similar to $g_{qg}^{(1)}(\lambda)$ in Eq.~(\ref{g1fun}) 
because $2 C_F \simeq C_A$. More precisely, since $2 C_F = C_A (1 - 1/N_c^2)$,
$g_{qq\to qq}^{(1)}(\lambda)$ is slightly smaller than $g_{qg}^{(1)}(\lambda)$
as long as they are evaluated in the perturbative region $\lambda = 
b_0\as\ln N < 1/2$.

We can now come back to the effect of the fragmentation component in $pN$
collisions. Using the resummed partonic cross sections in 
Eqs.~(\ref{gammaresqg}, \ref{gammaresf})
rather than their LO approximations, the right-hand side of 
Eq.~(\ref{qgqqratio}) has to be multiplied by an additional contribution,
as given by the ratio of the corresponding radiative factors, namely
\beq
\label{deltacorr}
\frac{\Delta_{N+1}^{qq \to qq}}{\Delta_{N+1}^{qg \to q\gamma}}
\simeq \frac{\exp \left\{ \left( C_F + 4C_F \right)
\frac{\as}{2\pi} \ln^2 N \right\}}{\exp \left\{ \left( 
C_F + 2C_A \right)\frac{\as}{2\pi} 
\ln^2 N \right\}} \;\;.
\eeq
Because of the relation $4 C_F \simeq 2C_A$ between the colour
charges, this factor does not sizeably differ from unity, as it can be
argued by its DL approximation on the right-hand side of
Eq.~(\ref{deltacorr}) (see also the comment below Eq.~(\ref{g1qqqq})).
We have thus shown that, at least at the LL level, soft-gluon
resummation does not enhance the relative importance of the
fragmentation component for prompt-photon production at large $x_T$.

The importance of the fragmentation component in $pN$ collisions
mainly depends on the detailed behaviour of the parton densities at
large $x$ and on how large are the values of $E_T$ of interest. This
issue, as well as the impact of the NLL corrections to the LL results
obtained above, require further studies that will be presented in a
future work. As for the present study, we limit ourselves to perform
soft-gluon resummation in the direct component and we check that the
fragmentation component does not sizeably contribute to the hadronic
cross section in the actual experimental configurations investigated
in the paper (see Sec.~\ref{datacomp}).

\subsection{Resummed cross section to NLL accuracy} 
\label{secresumxs}

We use soft-gluon resummation to NLL accuracy at the parton level
to introduce an improved prompt-photon cross section 
$\sigma_{\gamma, \,N}^{({\rm res})}(E_T)$ as follows
\beqn 
\sigma_{\gamma, \,N}^{({\rm res})}(E_T) &=& \sum_{ab=q{\bar q},\,qg\,,
{\bar q}g} 
f_{a/H_1,\,N+1}(\mu_F^2) \;f_{b/H_2,\,N+1}(\mu_F^2)
\nonumber \\
&\times& \left[ \;
{\hat \sigma}_{ab\to \gamma, \;N}^{({\rm res})}(\as(\mu^2); E_T^2, \mu^2, 
\mu_F^2) - \left( 
{\hat \sigma}_{ab\to \gamma, \;N}^{({\rm res})}(\as(\mu^2); E_T^2, \mu^2, 
\mu_F^2,) \right)_{\alpha \as^2} \, \right] \nn \\
\label{PPCSres}
&+& \sigma_{\gamma, \,N}^{({\rm NLO})}(E_T) \;\;,
\eeqn
where $\sigma_{\gamma, \,N}^{({\rm NLO})}$ is the prompt-photon 
hadronic cross section at NLO,
${\hat \sigma}_{ab\to \gamma, \;N}^{({\rm res})}$ is given in 
Eqs.~(\ref{gammaresqq})--(\ref{gammaresqbarg}) and
$\left( {\hat \sigma}_{ab\to \gamma, \;N}^{({\rm res})}\right)_{\alpha \as^2}$
represents its perturbative truncation at order $\alpha \as^2$ (i.e.\ at NLO).
Thus, because of the subtraction
in the square bracket on the right-hand side, Eq.~(\ref{PPCSres}) exactly
reproduces the NLO results and resums soft-gluon effects beyond
${\cal O}(\alpha\as^2)$ to NLL accuracy.
In general, we evaluate 
${\hat \sigma}_{ab\to \gamma, \;N}^{({\rm res})}$
using the NLL expression~(\ref{deltanll}) of the radiative factors
and including the ${\cal O}(\as)$ contribution~(\ref{cgamma}) of the constant
factors $C_{ab\to \gamma}(\as)$. This defines our NLO+NLL predictions.

The resummed formulae presented so far are given in $N$-moment space.
To obtain cross sections in the physical $x_T$-space (i.e.\ as functions
of the centre-of-mass energy), one has to perform
the inverse Mellin transformation:
\beq \label{MPPP}
E_T^3 \frac{d\sigma^{(\rm res)}_{\gamma}(x_T,E_T)}{dE_T} =\frac{1}{2\pi i}
\int_{C_{\rm MP}-i\infty}^{C_{\rm MP}+i\infty}\; dN \;x_T^{-2N}
\sigma_{\gamma, \,N}^{(\rm res)}(E_T) \;.        
\eeq

When the $N$-moments $\sigma_N$ are evaluated at a fixed perturbative order
in $\as$, they are analytic functions in a
right half-plane of the complex variable $N$. In this case, the constant 
$C_{\rm MP}$
that defines the integration contour in Eq.~(\ref{MPPP}) has to be chosen in
this half-plane, that is, on the right of all the possible singularities of 
the $N$-moments.

An additional complication occurs when the $N$-moments are computed in resummed
perturbation theory. In this case, since the resummed functions
$g_{ab}^{(1)}(\lambda)$ in Eq.~(\ref{g1fun}) (as well as the NLL functions
$g_{ab}^{(2)}$)
are singular at $\lambda=1/2$, the soft-gluon factors  
$\Delta_{N}(\as(\mu^2))$ in Eq.~(\ref{deltanll})
have cut singularities that start at the branch-point $N=N_L=\exp(1/2b_0\as)$. 
These singularities, which are related to the divergent behaviour of the 
perturbative running coupling  $\as$ near the Landau pole, signal the onset of
non-perturbative phenomena at very large values of $N$ or, equivalently,
in the region very close to threshold.  

The issue of how to deal with the Landau singularity in soft-gluon
resummation formulae for hadronic collisions was discussed in detail in 
Ref.~\cite{cmnt96}. In the evaluation of the inverse Mellin 
transformation~(\ref{MPPP}) we thus use the {\em Minimal Prescription}
introduced in  
Ref.~\cite{cmnt96}. The constant $C_{\rm MP}$ is chosen in such a way that all
singularities in the integrand are to the left of the integration contour,
except for the Landau singularity at $N=N_L$, that should lie to the far right.
This prescription is consistent~\cite{cmnt96} with the perturbative content of 
the soft-gluon resummation formulae because it converges asymptotically to the
perturbative series and it does not introduce (unjustified) power corrections
of non-perturbative origin. These corrections are certainly present in physical
cross-sections, but their effect is not expected to be sizeable
as long as $E_T$ is sufficiently perturbative and $x_T$ is
sufficiently far from the hadronic threshold. Obviously, approaching the
essentially non-perturbative regime $E_T \sim 1$~GeV, $x_T \to 1$,
a physically motivated treatment of non-perturbative effects has to be
introduced. In the following sections, we limit ourselves to presenting
numerical and phenomenological results that do not include any
non-perturbative correction.

\section{Results}
\label{Results}
We present in this section some numerical results, to provide an illustration
of the size of the effects considered and to show the improvements obtained
with respect to scale variations after the inclusion of the NLL corrections.

\subsection{Parton-level results}
\label{plres}
\begin{figure}
\begin{center}
\centerline{
\epsfig{file=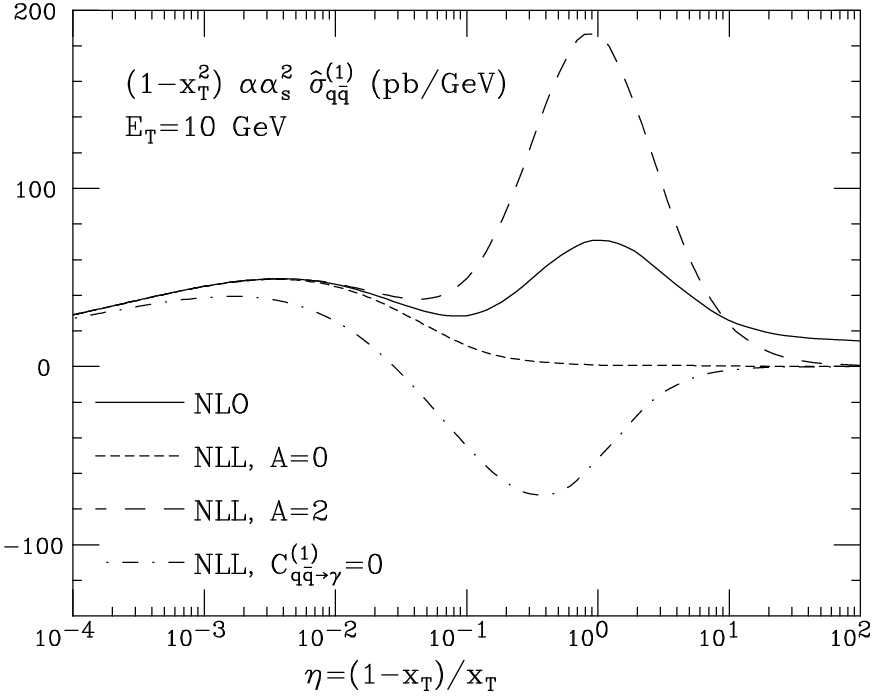,width=0.50\textwidth,clip=}\hfil
\epsfig{file=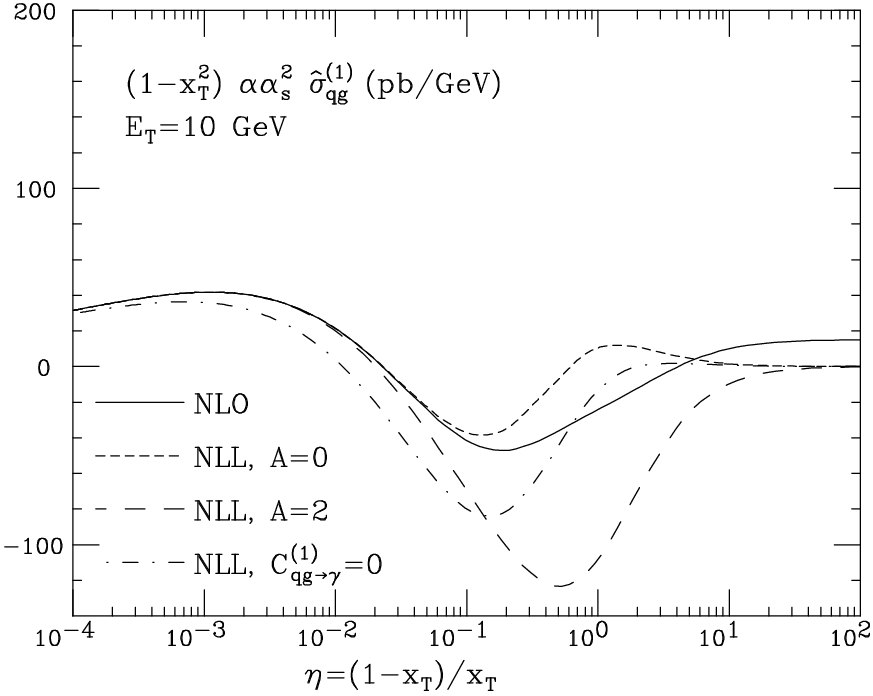,width=0.50\textwidth,clip=}                
}                                
\ccaption{}{\label{fig:qqnlo} 
Left (Right):
the \oaematwo\ contribution to the partonic prompt-photon cross section
for the process $q\bar q \to \gamma+X$ ($qg \to \gamma+X$)
plotted as a function of                      
$\eta=(1-\xt)/\xt$. The solid line represents the exact NLO
result of Ref.~\cite{Aurenche}; the short-dashed line 
is the \oaematwo\ 
piece of of the resummed result defined by                           
Eqs.~(\ref{gammaresqq}) and~(\ref{gammaresqg}); the dot-dashed line is
obtained from this last
result by setting the constant $C^{(1)}_{q\bar q \to \gamma}$ ($C^{(1)}_{qg
\to \gamma}$) to 0; 
the dashed line is obtained using
Eq.~(\ref{eq:Cqashift}) (Eq.~(\ref{eq:Cqgshift})), with $A=2$. The
renormalization, factorization and fragmentation scales were all set equal  
to $\mu^2=2\etsq$, and $e_q=1$.}
\end{center}
\end{figure}
     
We start by discussing the resummation effects at the level of partonic cross
sections. The resummed partonic cross section can be obtained from 
Eqs.~(\ref{PPCSres})
and (\ref{MPPP}) by assuming parton-density    
functions of the form $f_{a/H}(x)=\delta(1-x)$, and hence $f_{a/H, N}=1$ for 
all complex values of $N$. 

We consider first the $\oaematwo$ terms in the expansion of the resummed
cross section, in order to estimate to which accuracy this reproduces the
exact NLO results.                                                  
In Fig.~\ref{fig:qqnlo} (left) we plot the function  $\hat{\sigma}^{(1)}_{q\bar
q \to \gamma}/E_T^3$, defined in Eq.~(\ref{pxsg}), 
as a function of $\eta=(1-\xt)/\xt$.
The exact $\oaematwo$ result~\cite{Aurenche} is compared with  three possible
implementations of the resummation procedure, all equivalent at NLL accuracy. 
The first case (short-dashed line) corresponds to our default 
resummed prediction, as given by Eq.~(\ref{gammaresqq}).
In the second case (dot-dashed line), 
we set the constant $C_{q\bar q \to \gamma}^{(1)}$ introduced in
Eq.~(\ref{cgamma}) equal to 0.    
In the third case, we keep the contribution of the constant 
$C_{q\bar q\to \gamma}^{(1)}$, but
we modify it by a term suppressed by a factor of $1/N$, in order to explore
the possible effect of
contributions of order $1/N$ which cannot be taken into account by the 
soft-gluon resummation.
As a constraint on the form of these corrections, we must
impose that no poles appear on the positive real axis in the $N$ plane  
(these poles would logarithmically enhance the partonic cross section when
$\xt\to 0$).
We select a parametrization of the $1/N$ corrections that allows us to
bracket the exact result at \oaematwo\ :
\beq
 \label{eq:Cqashift}                            
  C_{q \bar q \to \gamma}^{(1)} \to C_{q \bar q \to \gamma}^{(1)} 
                                      \; (1+\frac{A}{N+A-1}) \;,
\eeq
\beq
\label{eq:Cqgshift}  
  C_{q g\to \gamma}^{(1)} \to C_{qg\to \gamma}^{(1)} 
                                      \; (1-\frac{A}{N+A-1}) \; , 
\eeq                                                
with $A>0$.
In our applications we shall consider the two cases with $A=0$ (namely no
correction to the $C_{ab\to \gamma}$ term) and $A=2$ as a way to establish the 
size of subleading threshold corrections beyond the NLL order.           
                               
\begin{figure}
\begin{center}
\centerline{
\epsfig{file=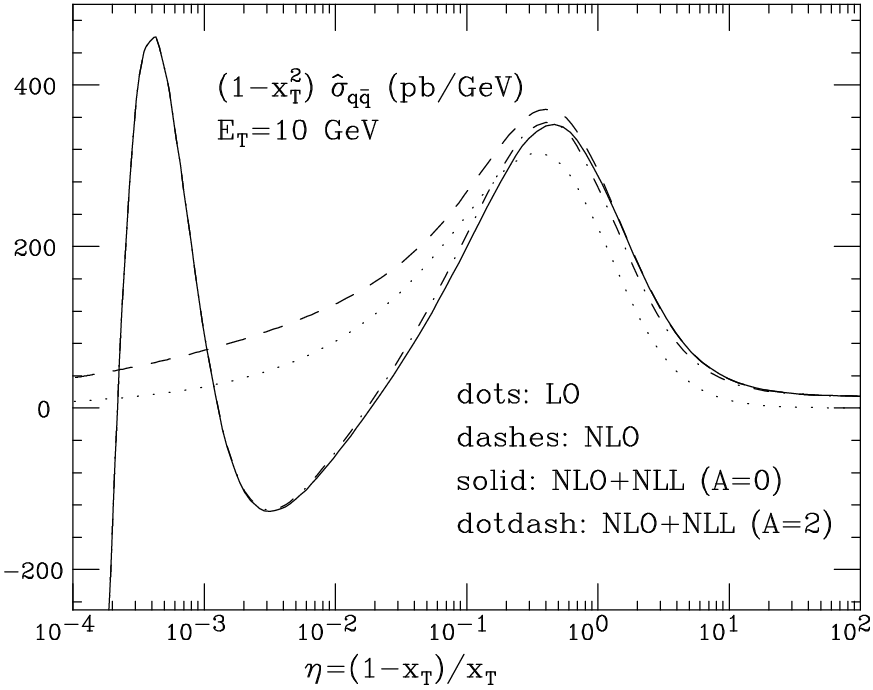,width=0.50\textwidth,clip=}\hfil
\epsfig{file=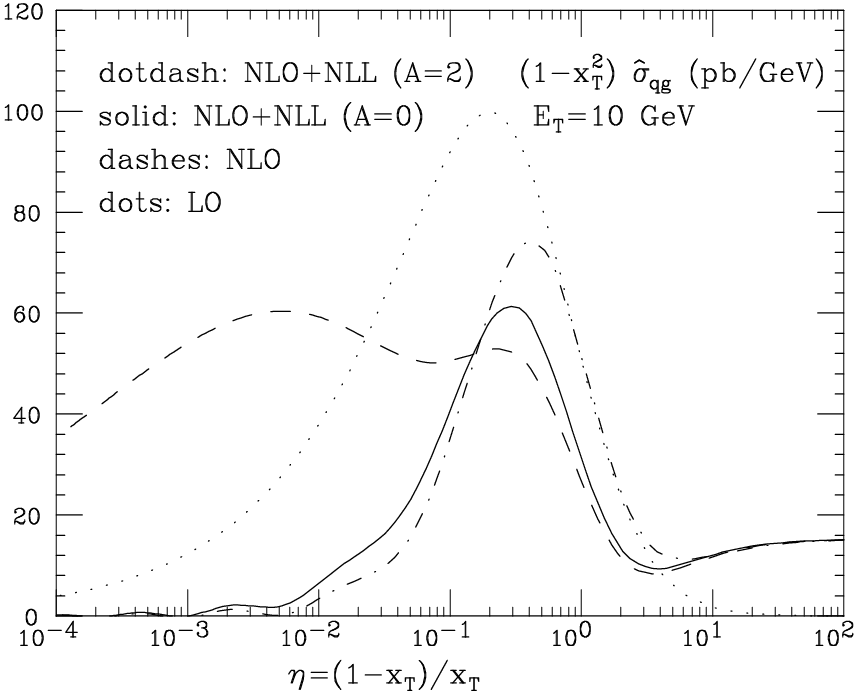,width=0.50\textwidth,clip=}
}                                
\ccaption{}{\label{fig:qares10} 
Partonic cross-section for the processes $q\bar q \to \gamma X$  (left)
and $q g \to \gamma X$  
(right) (in pb/GeV,                                                    
and for $\et=10$~GeV). The dotted line is the LO result;
the dashed line is the exact NLO
result;  the solid (dotdashed) lines correspond to the NLO+NLL
result, with the coefficient $A$ defined in Eq.~(\ref{eq:Cqashift}) (left)
and in Eq.~(\ref{eq:Cqgshift}) (right) equal to 0 (2). The number of flavours
$N_f$ was set equal to 4 and we have taken $\LambdaQCD^{(4)}=0.151$~GeV.}
\end{center}
\end{figure}

\begin{figure}
\begin{center}
\centerline{
\epsfig{file=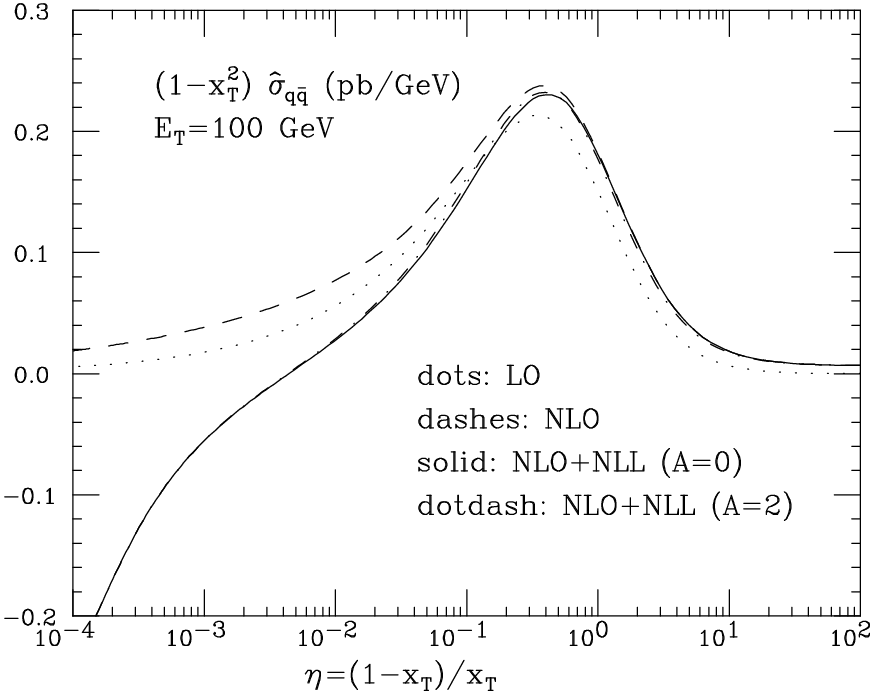,width=0.50\textwidth,clip=}\hfil
\epsfig{file=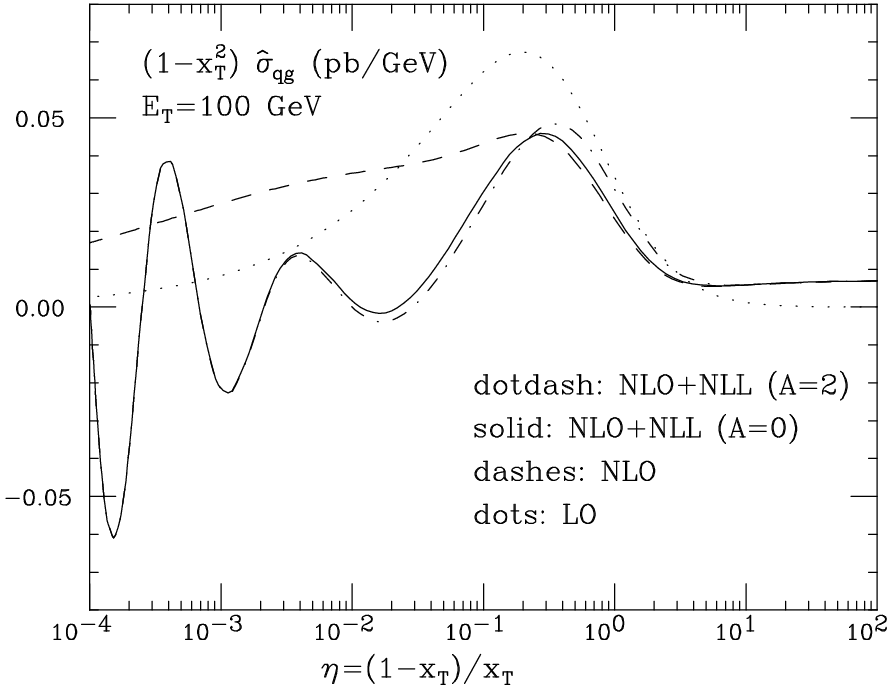,width=0.50\textwidth,clip=}
}                                
\ccaption{}{\label{fig:qares100} 
Same as Fig.~\ref{fig:qares10}, for 
$\et=100$~GeV. }
\end{center}   
\end{figure}

As one can see from Fig.~\ref{fig:qqnlo}, the inclusion of the finite
term $C_{q\bar{q}\to \gamma}^{(1)}$ is essential to accurately
reproduce shape and normalization of the exact \oaematwo\ result not
only near threshold, but below it as well.  The agreement deteriorates
unavoidably for $\eta \gg 1$, as, here, terms subleading in $1/N$
become important.
      
Analogous results for the $qg$ channel are given in the right panel of
Fig.~\ref{fig:qqnlo}.  Note that in both cases the two choices $A=0$
and $A=2$ in Eq.~(\ref{eq:Cqgshift}) bracket the exact result over a
large region of $\eta$, and thus provide a good estimator of the
subleading terms' systematics.  The choice $A=0$, furthermore,
provides a very accurate description up to values of $\eta$ of the
order of 1/10.

The fully-resummed parton-level cross sections are shown in
Fig.~\ref{fig:qares10}  for the $q\bar q$ and $qg$
channels (left and right panel, respectively), and for $E_T=10$~GeV
(Fig.~\ref{fig:qares100} collects the same results for $E_T=100$~GeV).  
Here and in the following we shall                        
define the resummed cross sections as in Eq.~(\ref{PPCSres}), that is, we
substitute their $\oaematwo$ terms with                                 
the exact NLO result,   
using the same choice of renormalization and factorization scales.
In this way our results are exact up to (and including)
$\oaematwo$, and include the NLL resummation of terms of $\oaemacube$ and 
higher.
We compare the fixed-order results (dashed lines) with the resummed results.
For these we provide both the $A=0$ and $A=2$ prescriptions.
Note that, even at the level of resummed cross sections, the      
difference between the $A=0$ and $A=2$ results are rather small, in particular
for the $q \bar q$ channel.
   
\begin{figure}
\begin{center}
\centerline{
  \epsfig{file=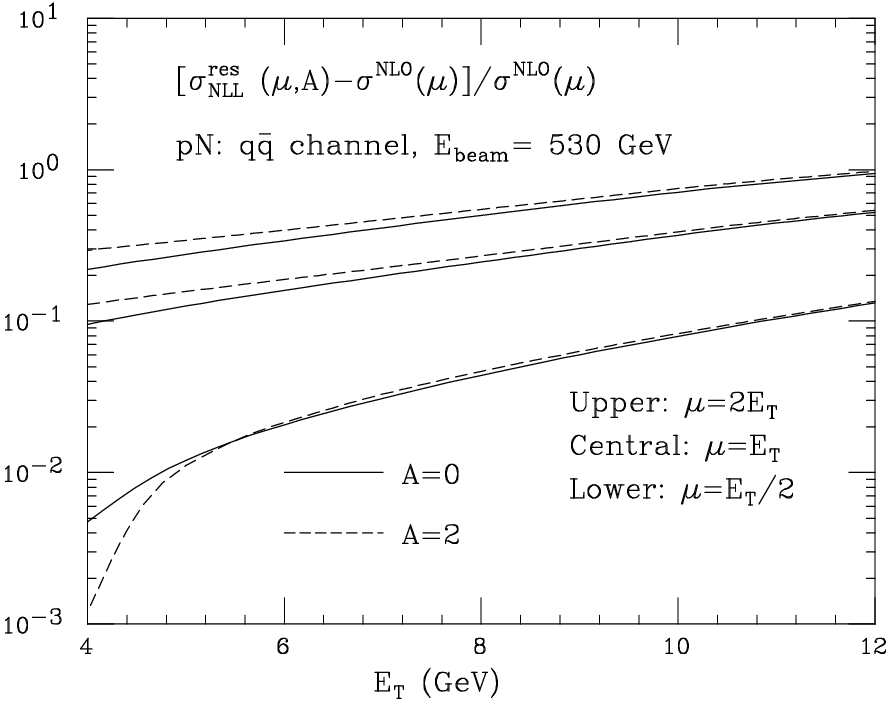,width=0.5\textwidth,clip=}
  \epsfig{file=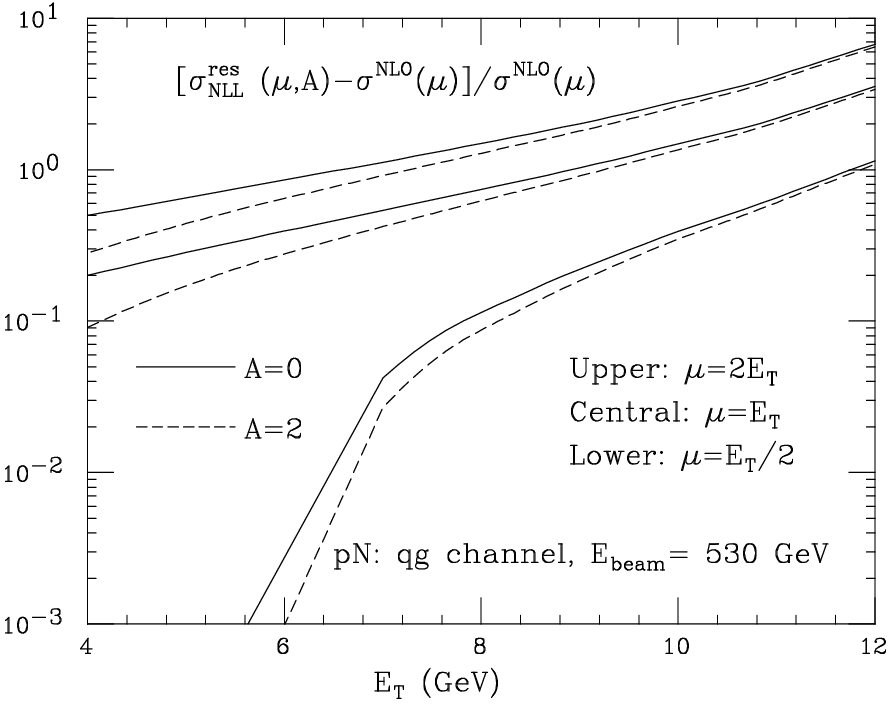,width=0.5\textwidth,clip=}}
\vspace*{\mysep}
\ccaption{}{\label{fig:deltaqa} 
Contribution of gluon resummation at order $\oaemacube$ and higher, relative to
the exact NLO result, for photon production via $q\bar q$ (left plot)
and $qg$ (right plot) annihilation, in
$pN$ collisions at $\sqrt{S}=31.5$~GeV ($E_{\rm beam}=530$~GeV).
The solid (dashed) lines correspond to $A=0$ ($A=2$).      
The three sets of curves correspond to the choice of scale $\mu=\mu_F=2\et,\;
\et$ and $\et/2$, in descending order, with
PDF set CTEQ4M and $N_f=5$.}         
\end{center}                   
\vspace*{\mysep}
\end{figure}

\begin{figure}
\begin{center}
\epsfig{file=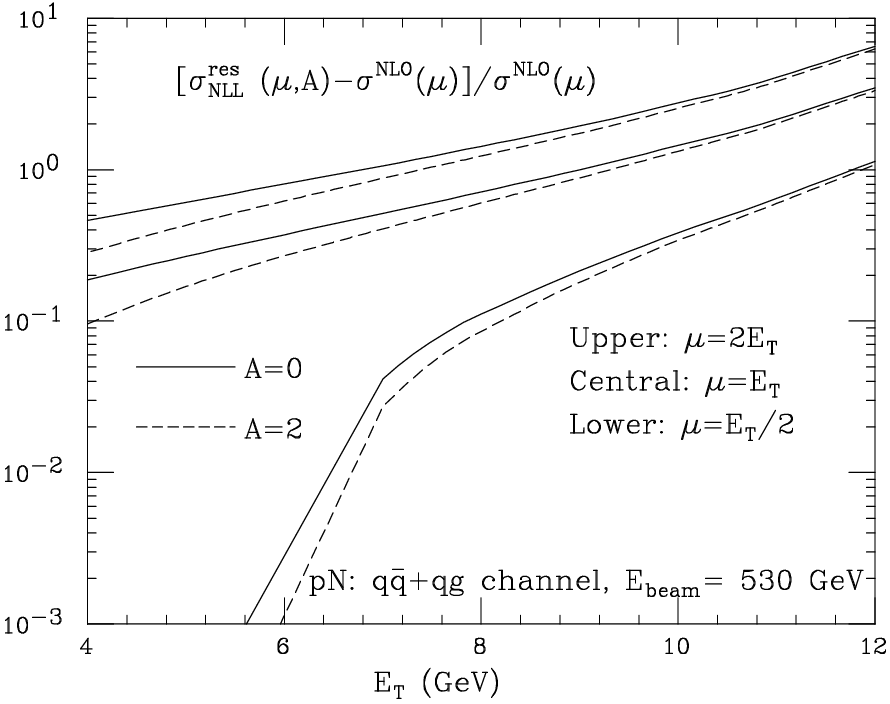,width=0.7\textwidth,clip=}
\vspace*{\hmysep}     
\ccaption{}{\label{fig:deltatot} 
Same as Fig.~\ref{fig:deltaqa}, for 
the combined production channels $qg+q\bar q$.}
\end{center}
\end{figure}

\subsection{Hadron-level results}
\label{HLResults}
In this section we present some results for the full hadronic cross sections.
The main points we intend to highlight are:
\begin{enumerate}
\item the size of the NLL corrections, relative to the NLO contributions;
\item the scale dependence at NLL order.
\end{enumerate}
Our goal here is to explore the pure effects of
resummation at higher orders. Therefore we shall neglect in this
section all production channels which are not improved by the resummation
corrections considered in this work. This includes all processes which
first appear at \oaematwo, such as $gg\to q\bar q\gamma$ and $qq^{(\prime)}\to
qq^{(\prime)}\gamma$,  as well as
all contributions proportional to a parton  $\to\gamma$ fragmentation
function, as discussed in the previous section.
These terms will however be included, at fixed NLO, in our comparison with
experimental data, performed in the next section.

To be more specific, we list here the classes of diagrams included at
NLO, in addition to the LO processes $q\bar q \to g\gamma$ and
${q} g\to {q} \gamma$ (the possible replacement of quarks with
antiquarks in all these cases is understood):
\be
\label{eq:procsin}
{q} g\to {q} g \gamma \; ,\quad
q\bar q \to q\bar q \gamma \; ,\quad
q\bar q \to q'\bar q' \gamma \; ,\quad
q\bar q \to gg \gamma \; .
\ee
For the third set we only include the diagrams proportional to
$e_q^2$, since the part of the amplitude describing the photon
emission from the final-state quarks (which is by itself gauge
invariant), cannot be considered as a correction to any tree-level process.

As a default set of parton densities we
shall use the CTEQ4M set described in Ref.~\cite{cteq4}. For the
purposes of the present study, no significant change is obtained
if different sets are used.

Figures~\ref{fig:deltaqa} and~\ref{fig:deltatot} present the ratios:
\beq \label{deltadef} 
    \frac{\sigma_{{\rm NLL,\ q\bar{q}}}^{\rm res} 
         - \sigma_{{\rm q\bar{q}}}^{\rm NLO}}
         {\sigma_{{\rm q\bar{q}}}^{\rm NLO}}
    \,,\quad 
    \frac{\sigma_{{\rm NLL,\ qg}}^{\rm res} - \sigma_{{\rm qg}}^{\rm NLO}}
         {\sigma_{{\rm qg}}^{\rm NLO}}
    \,,\quad                   
    \frac{\sigma_{{\rm NLL,\ (qg+q\bar{q})}}^{\rm res} 
         - \sigma_{{\rm (qg+q\bar{q})}}^{\rm NLO}}
         {\sigma_{{\rm (qg+q\bar{q})}}^{\rm NLO}}
  \, ,
\eeq
where, for simplicity, we indicated here with $\sigma$ the differential
distribution $d\sigma/dE_T$.
For each channel we present the results using both the $A=0$ (solid lines) and
$A=2$ (dashed lines) prescriptions. We also show the dependence on
the choice of renormalization and factorization scales, which we take
equal, and varying within the set $\mu=\mu_F=(\et/2,\,\et,\,2\et)$. In
this section we shall always keep the fragmentation scale $\mu_f$,
necessary for the factorization of the singularities from final-state
collinear photon emission, equal to \et.
Note that the size of the resummation effects is larger for the larger
scales, contrary to the behaviour of the scale dependence of the NLO cross
section. This suggests that the scale dependence of the resummed cross section
will be reduced relative to that of the NLO results. 
 
\begin{figure}
\begin{center}
\epsfig{file=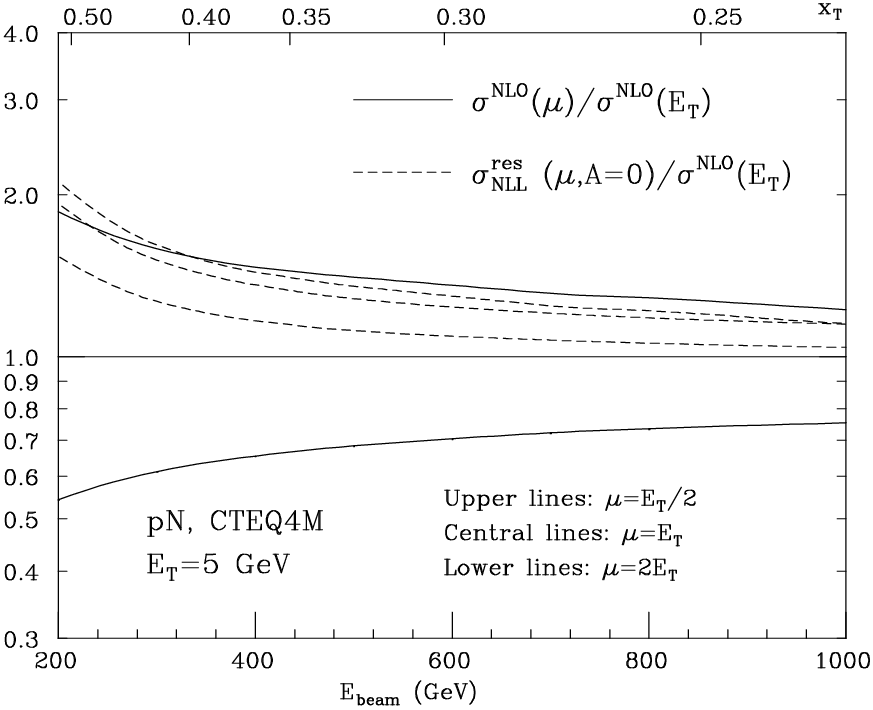,width=0.7\textwidth,clip=}
\ccaption{}{\label{fig:scdepl0} 
    Scale dependence of $d\sigma/dE_T$ ($qg+q\bar{q}$ components) for
  prompt photons in $pN$ collisions, at $E_T=5$~GeV, plotted as a
  function of the proton-beam energy, $E_{\rm beam}$ (the associated
  values of \xt\ are given on the top scale).  The solid lines
  represent the exact NLO result for different choices of $\mu=\mu_F$
  ($\mu=\et/2$ and $2\et$), normalized to the $\mu=\et$ result.  The
  dashed lines represent the NLO+NLL result (with $A=0$)
  for different choices of $\mu$, normalized to the NLO
  $\mu=\et$ result.}
\end{center}                                                              
\vspace*{\hmysep}
\end{figure}

\begin{figure}
\begin{center}
  \centerline
  {\epsfig{file=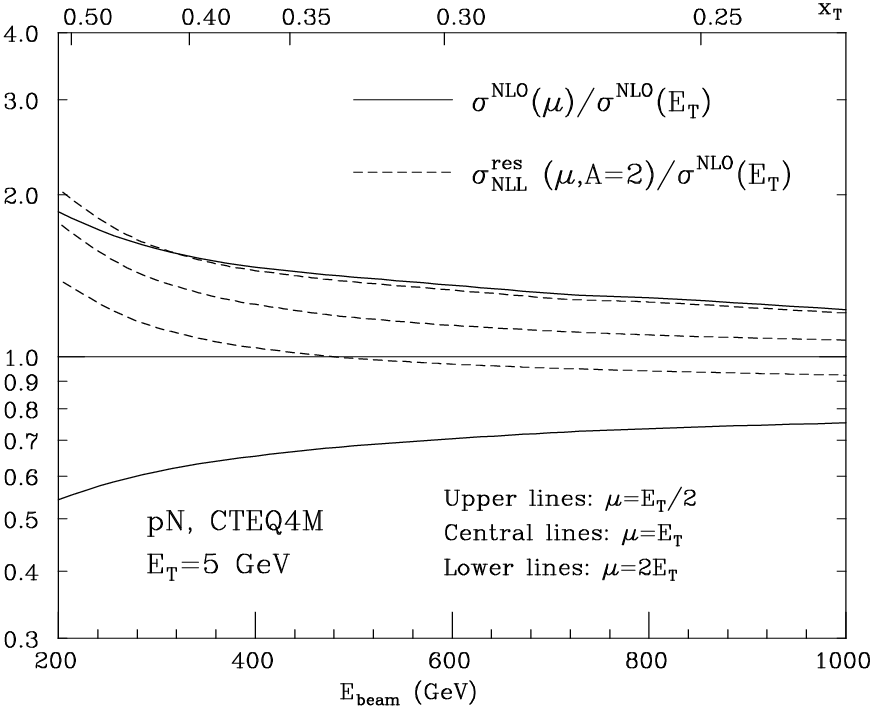,width=0.5\textwidth,clip=}
   \epsfig{file=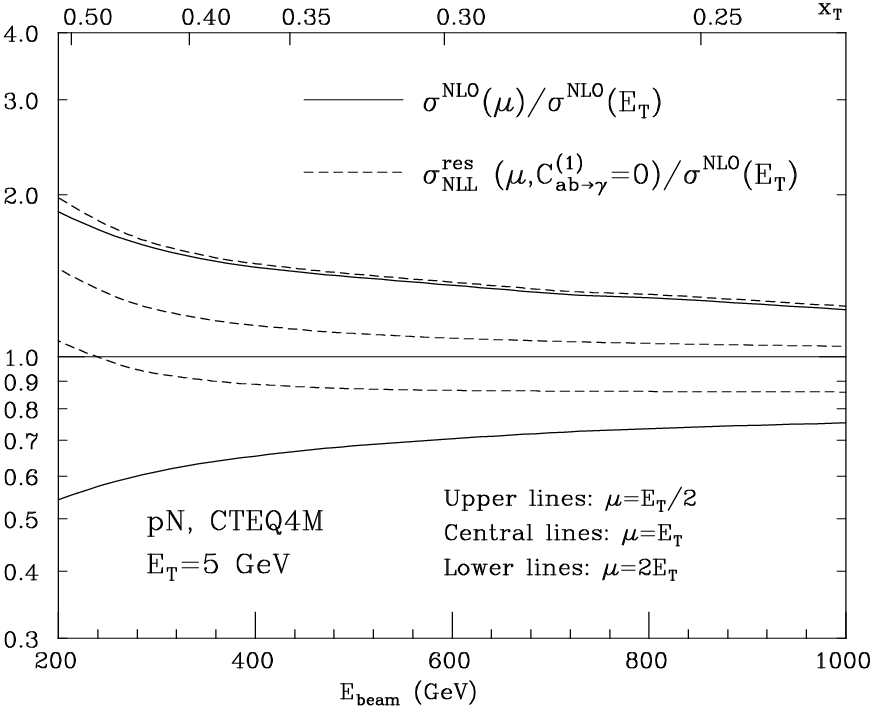,width=0.5\textwidth,clip=}}
  \ccaption{}{\label{fig:scdepl} Same as Fig.~\ref{fig:scdepl0}, but
    with $A=2$ (left panel) and 
    with $C_{ab\to\gamma}^{(1)}=0$ (right panel). }
\end{center}                                  
\end{figure}

The scale dependence of the resummed cross section ($qg+q\bar{q}$
contributions), compared to the NLO one, is given in Fig.~\ref{fig:scdepl0},
for the $A=0$ case.
The same result for $A=2$ is given in the left panel of
Fig.~\ref{fig:scdepl}.
We plot the distributions as a function of the beam energy
($E_{\rm beam}$) for the fixed value of $\et=5$~GeV. Different values of
$E_{\rm beam}$, therefore, probe different ranges of \xt, as indicated by the
upper labels on the plots.

Note the significant reduction in scale dependence, more marked in the $A=0$
case. To display the importance of the inclusion of the 
constant $C_{ab \to \gamma}^{(1)}$ terms, we show the same
scale-dependence plot with $C_{ab \to \gamma}^{(1)}=0$ in the right
panel of Fig.~\ref{fig:scdepl}. While the scale sensitivity is
slightly worse than in the cases with $C_{ab \to \gamma}^{(1)}\ne 0$,
there is still an important improvement over the NLO behaviour.
                                        
\begin{figure}                                      
\begin{center}
  \epsfig{file=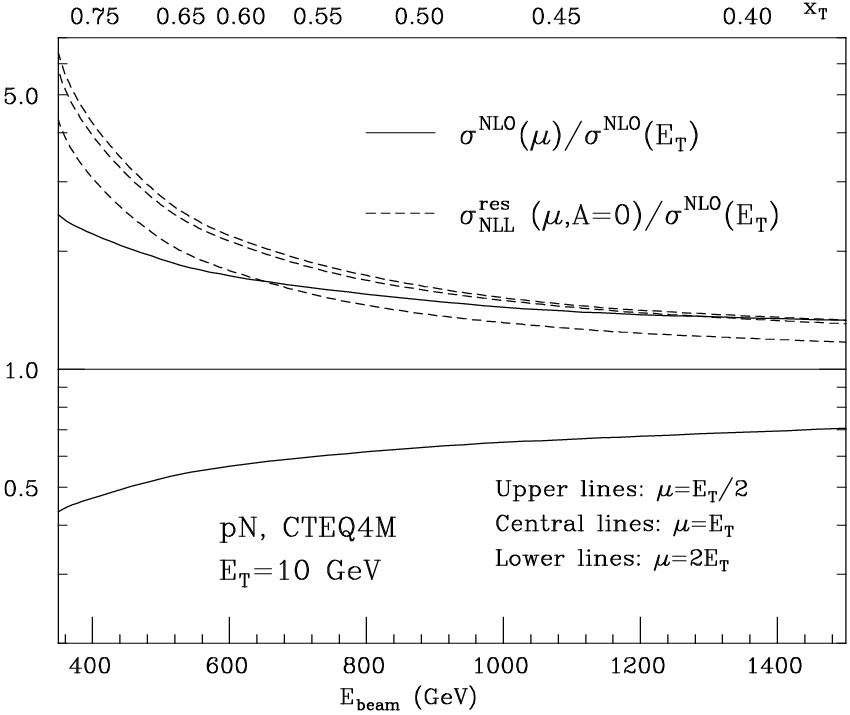,width=0.7\textwidth,clip=}
  \ccaption{}{\label{fig:scdeph0} Scale dependence of $d\sigma/dE_T$
    ($qg+q\bar{q}$ components) for prompt photons in $pN$ collisions,
    at $E_T=10$~GeV, plotted as a function of the proton-beam energy,
    $E_{\rm beam}$ (the associated values of \xt\ are given on the top
    scale).  The solid lines represent the exact NLO result for
    different choices of $\mu=\mu_F$ ($\mu=\et/2$ and $2\et$),
    normalized to the $\mu=\et$ result.  The dashed lines represent
    the NLO+NLL result (with $A=0$) for
    different choices of $\mu$, normalized to the NLO $\mu=\et$
    result.}
\end{center}                                                              
\vspace*{\hmysep}
\end{figure}

\begin{figure}
\begin{center}
  \centerline{ \epsfig{file=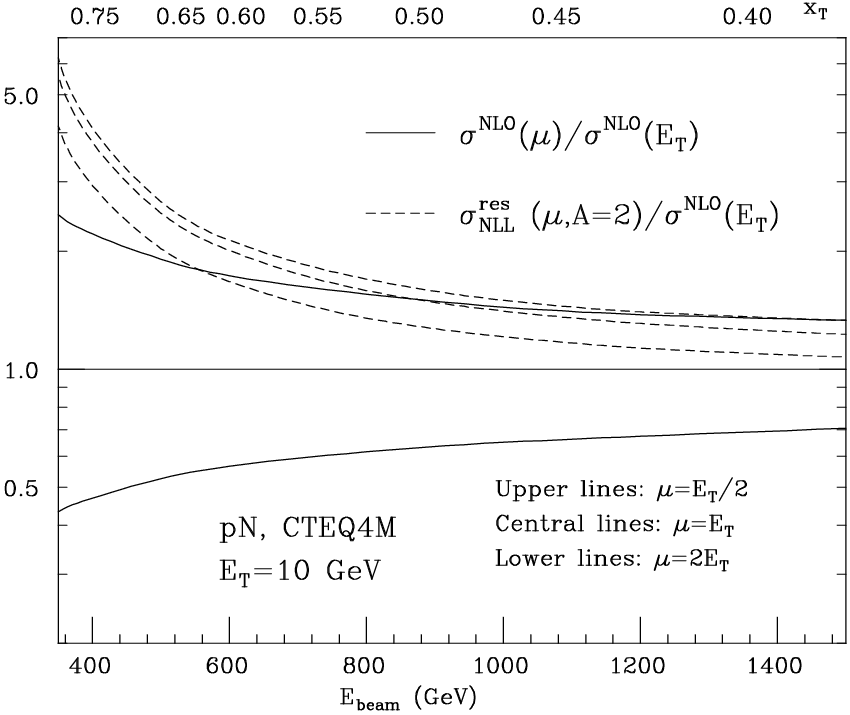,width=0.5\textwidth,clip=}
    \epsfig{file=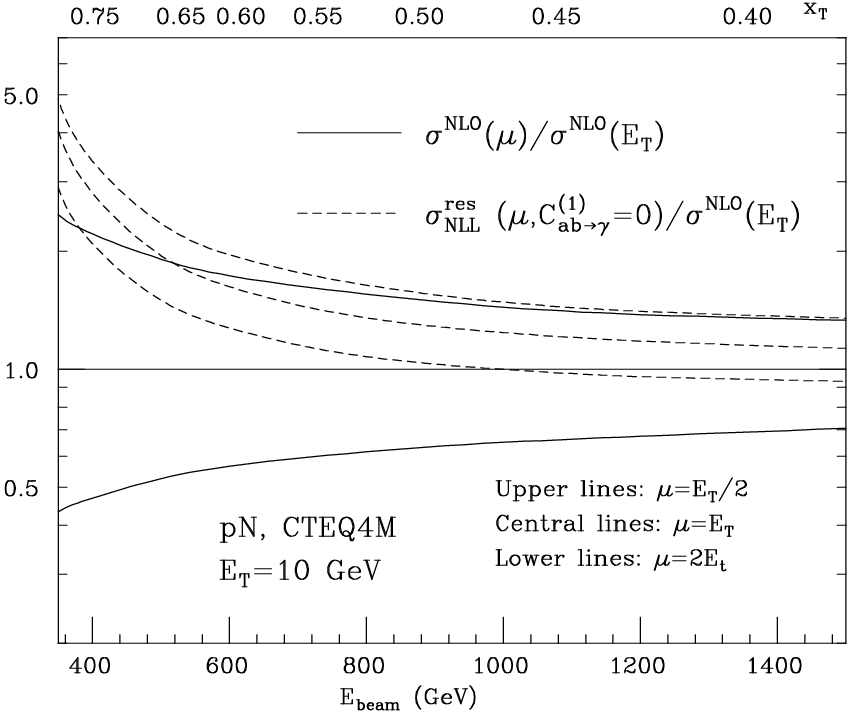,width=0.5\textwidth,clip=} }
  \vspace*{\mysep} \ccaption{}{\label{fig:scdeph} Same as
    Fig.~\ref{fig:scdeph0}, but with with $A=2$ (left panel) and with
    $C_{ab\to\gamma}^{(1)}=0$ (right panel). }
\end{center}                                  
\end{figure}

Similar results, for $\et=10$~GeV, are shown in 
Figs.~\ref{fig:scdeph0} and~\ref{fig:scdeph}. The general features of these
distributions are similar to those of the plots for $\et=5$~GeV. Small
violations of \xt-scaling can be observed between $\et=5$ and $10$~GeV, due to
the evolution of the coupling constant and of parton densities.

\begin{figure}
\begin{center}
\epsfig{file=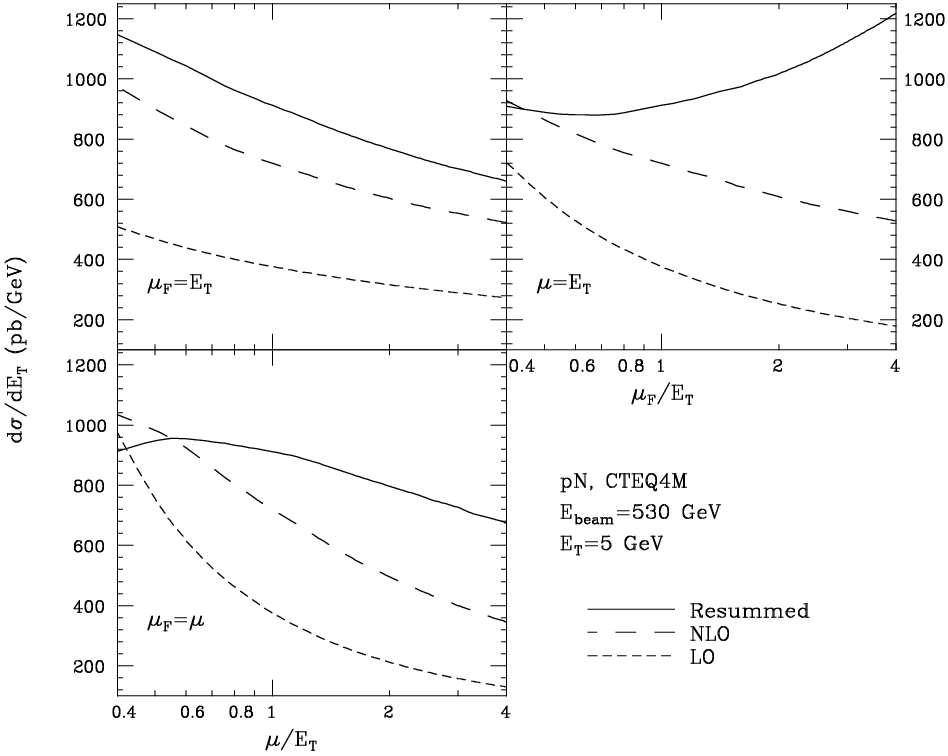,width=0.8\textwidth,clip=}
\vspace*{\mysep}                 
\ccaption{}{\label{fig:scdep5} 
Scale dependence of the differential $E_T$ distribution in $pN$
collisions, for $E_T=5$~GeV. 
We compare the results at the Born and NLO level with the results
of the resummed calculation.
Upper left: 
renormalization-scale dependence, 
with the factorization scale fixed to 
$\muf=E_T$. Upper right: 
factorization-scale dependence, 
with the renormalization scale fixed to 
$\mu=E_T$. Lower left:
scale dependence, with the renormalization and factorization scales equal.}
\end{center}                                                        
\end{figure}

\begin{figure}
\begin{center}
\epsfig{file=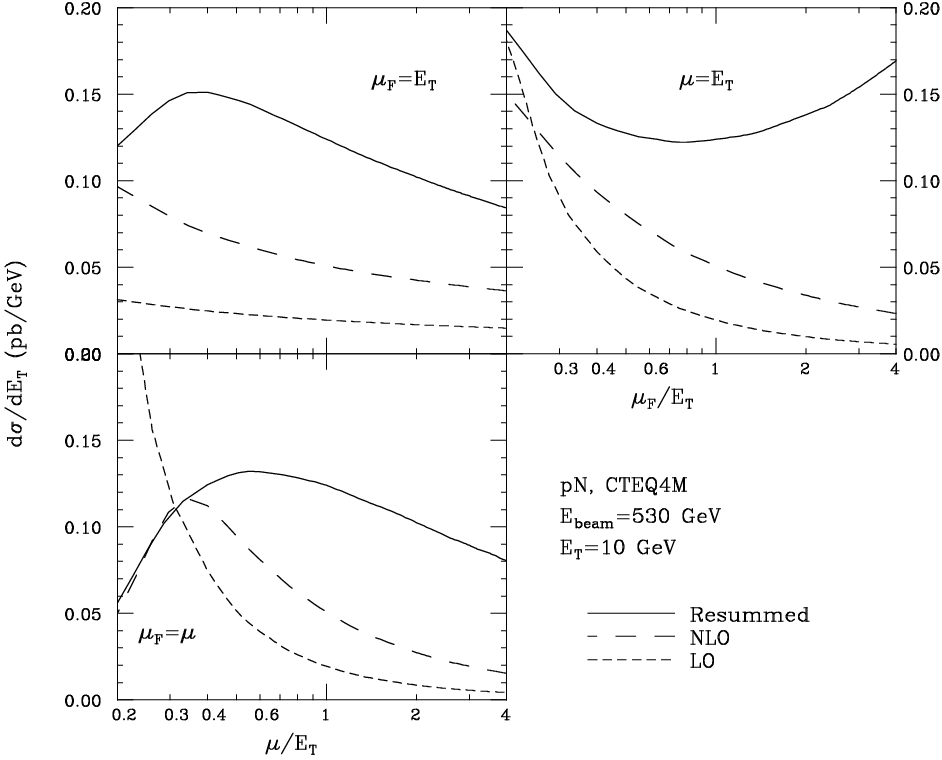,width=0.8\textwidth,clip=}
\vspace*{\mysep}      
\ccaption{}{\label{fig:scdep10} 
Same as Fig.~\ref{fig:scdep5}, for $E_T=10$~GeV. }
\end{center}                                  
\end{figure}

We also explored the independent renormalization- and factorization-scale
dependence of our calculations. The large size of this dependence at
NLO was stressed already in Refs.~\cite{Vogelsang95,Aurenche98}.
The results, for $pN$ collisions at $E_{\rm beam}=5$
and 10~GeV, are shown in Figs.~\ref{fig:scdep5} and \ref{fig:scdep10},
respectively. With the exception of the renormalization-scale dependence at 
5~GeV, a significant improvement in the stability of the results,
relative to the dependence at LO and NLO, is observed in all cases. 

{\renewcommand{\arraystretch}{1.8}
\begin{table}
\begin{center}
\begin{tabular}{cccccccc} \hline
Channel & $E_T$ (GeV)& $\alpha\alpha_s$ & $\alpha\alpha_s^2$ & $\alpha\alpha_s^3$ &
$\alpha\alpha_s^4$ &
$\alpha\alpha_s^5$ & $\alpha\alpha_s^{\ge 6}$  \\ \hline \hline
$q g$ (pb/GeV) & 5 &   329 &   325 &   126 & 39 &   11 &   6\\
$q\bar q$ (pb/GeV) & 5& 47 &   29 &   7.3 & 1.7 & 0.32 & 0.053 \\
$qg$ (fb/GeV) & 10 & 19 &  31 &  28 & 20 & 12 & 13 \\         
$q\bar q$ (fb/GeV) & 10 & 0.83 &  0.73 & 0.36&  0.15 & 0.05& 0.02 \\
\hline
\end{tabular}                                                                 
\ccaption{}{\label{tab:toppert} Contributions to the prompt-photon 
rate $d\sigma/d\et$ in $pN$ collisions at $E_{\rm beam}=530$~GeV,
from higher orders in the expansion of the NLL resummed result. 
Results for $\et=5$ (10)~GeV are shown in the
first (second) two rows. The renormalization, factorization and
fragmentation scales 
are set equal to $\mu=\muf=\mu_f=\et$, and the PDF set is CTEQ4M.    
The $\alpha\alpha_s^2$ column gives the {\em exact} NLO result. }
\end{center}                                         
\end{table} }
The convergence of the higher-order corrections is displayed in
Table~\ref{tab:toppert}. 
The last column includes the sum of all contributions of order            
$\alpha\alpha_s^6$ and higher, performed using the {\em Minimal Prescription}
of Eq.~(\ref{MPPP}). 
The fixed-order terms do not have any ambiguity due to the choice
of the contour for the Mellin transformation in Eq.~(\ref{MPPP}).
The contribution from the $q\bar q$ channel converges very rapidly. In the case
of the $qg$ channel the convergence is slower, in particular at the larger
values of $x_T$, but even at $\et =10$~GeV the size of the resummed
contributions 
beyond order $\alpha\alpha_s^6$ is only of the order of 10\% of the total.
This supports the validity of the {\em Minimal Prescription},
since the truncated resummed expansion converges to it very smoothly.
                             
\section{Comparison with current data}
\label{datacomp}
We present in this section a comparison between our calculations and
the results of the two most recent measurements of fixed-target
direct-photon production. 
E706~\cite{Huston98} studied photons produced in $p{\rm Be}$ collisions at
$E_{\rm beam}=530$~GeV, covering the centre-of-mass rapidity range
$\vert y \vert <0.75$\footnote{In this section we shall use $y$ to
  indicate the value of photon rapidity in 
  the hadron-hadron centre-of-mass frame.}. 
The data span the \et\ region between 3.5
and 11~GeV, approximately. This corresponds to $0.22<\xt<0.70$. 
UA6~\cite{UA6} studied photons produced in ${\bar p} p$ collisions at
$E_{\rm beam}=315$~GeV, over the rapidity range $-0.1< y<0.9$. The
\et\ region extends between 4 and 7~GeV ($0.33 <\xt< 0.58$), approximately.
What follows is not meant to be a systematic
phenomenological study, but will serve as a benchmark to assess the impact
of the resummation effects in realistic experimental conditions.
All the calculations in this section have been done assuming
$\mu=\muf=\mu_f$.
For recent complete studies of all available data, done using the fixed-order
NL calculations, see Refs.~\cite{Huston99,Aurenche98,Martin98}.

To compare our resummed predictions with actual data, two additional
things need to be done: inclusion of the $1/N$-suppressed
contributions, and inclusion of realistic experimental cuts.  

\begin{figure}
\begin{center}
\centerline{
\epsfig{file=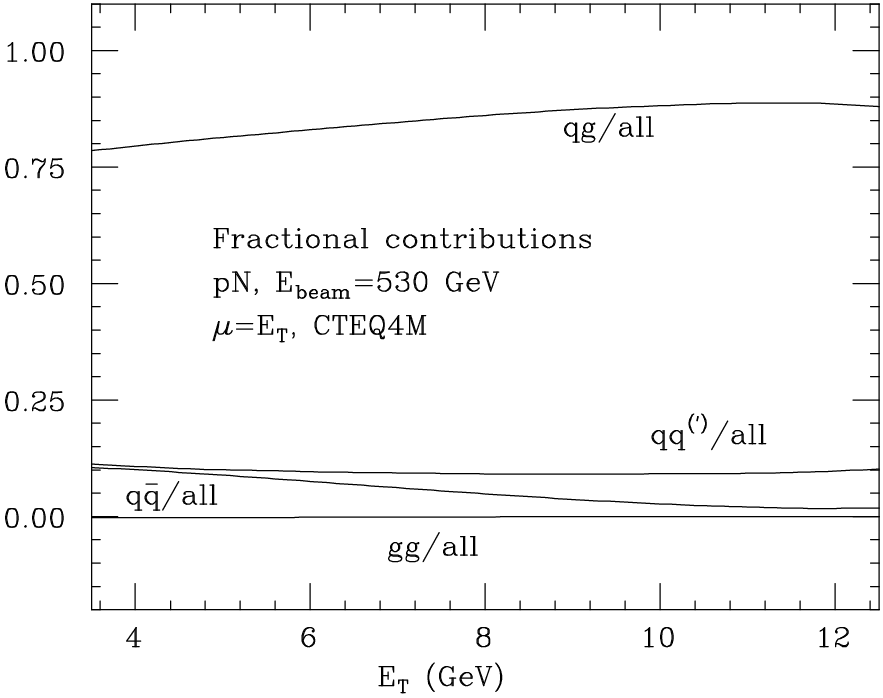,width=0.5\textwidth,clip=}
\epsfig{file=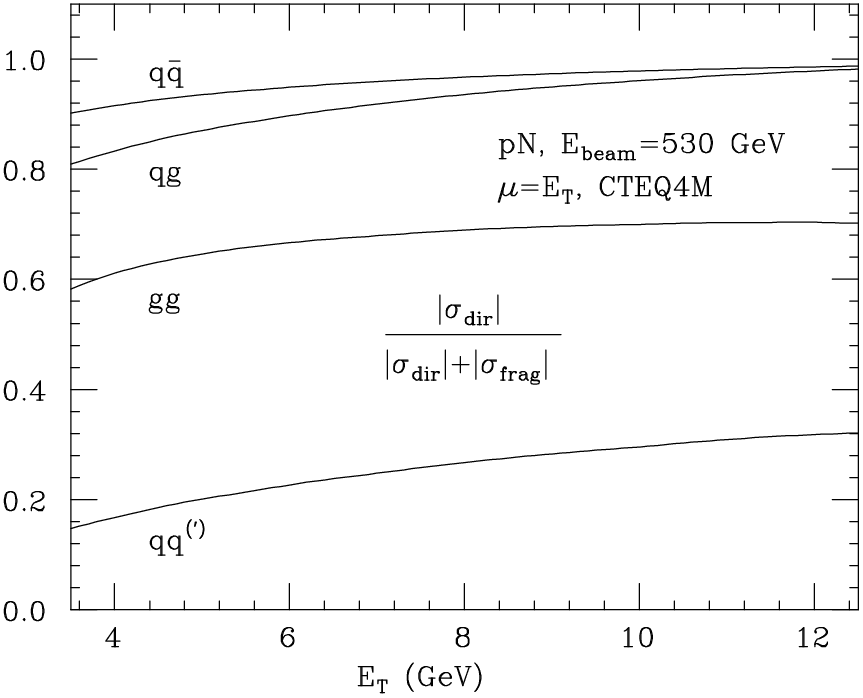,width=0.5\textwidth,clip=}
}
\ccaption{}{\label{fig:e706frag} 
Left: relative contribution of different initial states to the
$\et$ distribution in $pN$ collisions at $E_{\rm beam}=530$~GeV.
Right: relative size of the direct contribution vs. the 
sum of direct and fragmentation one, for the different channels.}
\end{center}                                                              
\end{figure}

\begin{figure}
\begin{center}
\centerline{
\epsfig{file=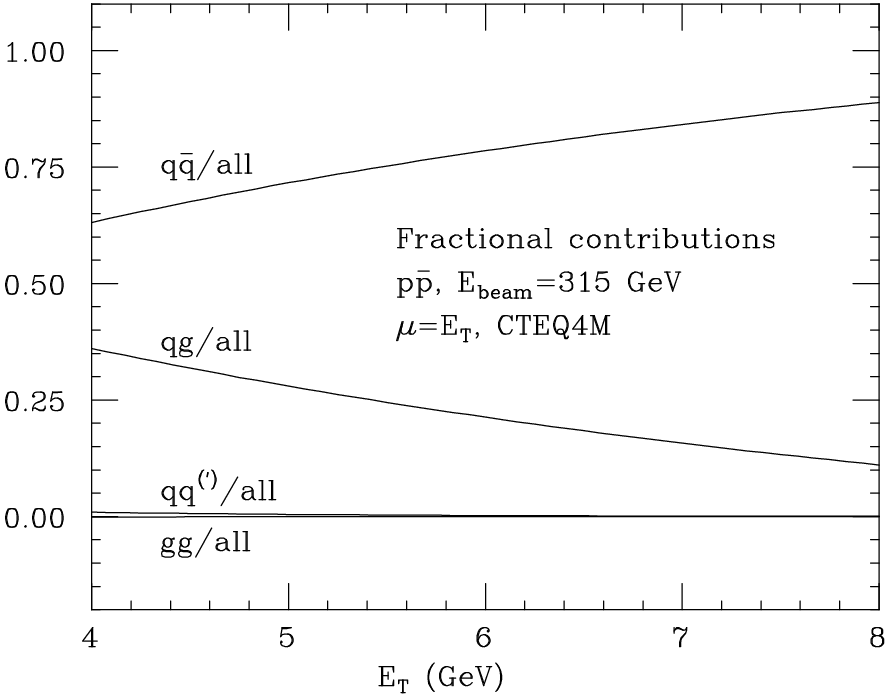,width=0.5\textwidth,clip=}
\epsfig{file=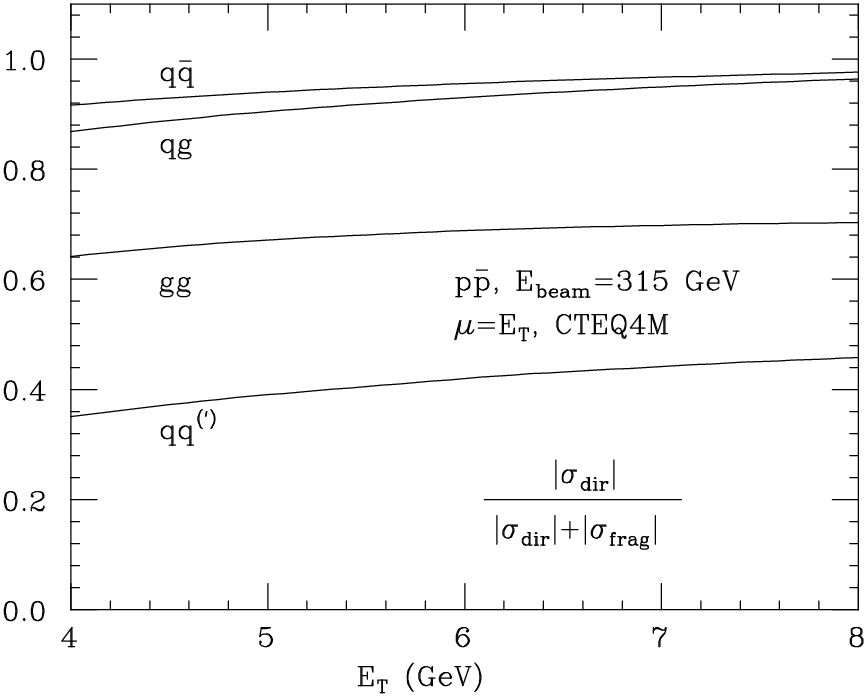,width=0.5\textwidth,clip=}
}
\ccaption{}{\label{fig:ua6frag} 
Left: relative contribution of different initial states to the
$\et$ distribution in ${\bar p} p$ collisions at $E_{\rm beam}=315$~GeV.
Right: relative size of the direct contribution vs. the 
sum of direct and fragmentation one, for the different channels.}
\end{center}                                                              
\end{figure}

The
class of processes for which we evaluated resummation corrections in
section~\ref{secnotat} provides the dominant contribution to the
production rate in realistic experimental configurations. On the
left-hand side of 
Fig.~\ref{fig:e706frag} we plot the relative contribution, evaluated at NLO, of
the processes with a $qg$, $q\bar q$, $qq^{(\prime)}$ and $gg$ initial
state for the E706 experimental configuration. As one can see, the
sum of $qg$ and $q\bar q$ accounts for 90\% of the overall rate,
independently of $\et$. On the right-hand side of
Fig.~\ref{fig:e706frag} we plot the
rate of the direct contributions, relative to the sum of direct and
fragmentation, for each given channel. In the case of the
$qq^{(\prime)}$ and $gg$ channels we compare the absolute values of
the rates, since the direct component is negative after the
subtraction of the initial-state mass singularities. The comparison of
the two plots in Fig.~\ref{fig:e706frag} shows that the processes
for which we are going to include resummation corrections account at NLO for a
fraction of the total rate between 70 and 90\%, in the \et\ range
4--12~GeV. The situation is even better in ${\bar p} p$
collisions (see Fig.~\ref{fig:ua6frag}, obtained for the UA6
experimental configuration), where the to-be-resummed
processes account for over 90\% of the NLO rate.

We therefore expect that the neglect of the resummation corrections to
the $gg+qq^{(\prime)}$ and to the fragmentation processes is only a
minor correction to the overall picture. For the present study, and in
addition to the resummed predictions for the processes listed in
Eq.~(\ref{eq:procsin}), we will therefore only include the fixed-order
NL determination of these remaining components of the direct-photon
production process. The fragmentation processes are evaluated using
the NLO single-inclusive parton \et\ distributions from~\cite{ACGG},
convoluted with the GRV photon fragmentation functions~\cite{GRV}. We
found very small sensitivity to the choice of the photon fragmentation
functions.

\begin{figure}
\begin{center}
\epsfig{file=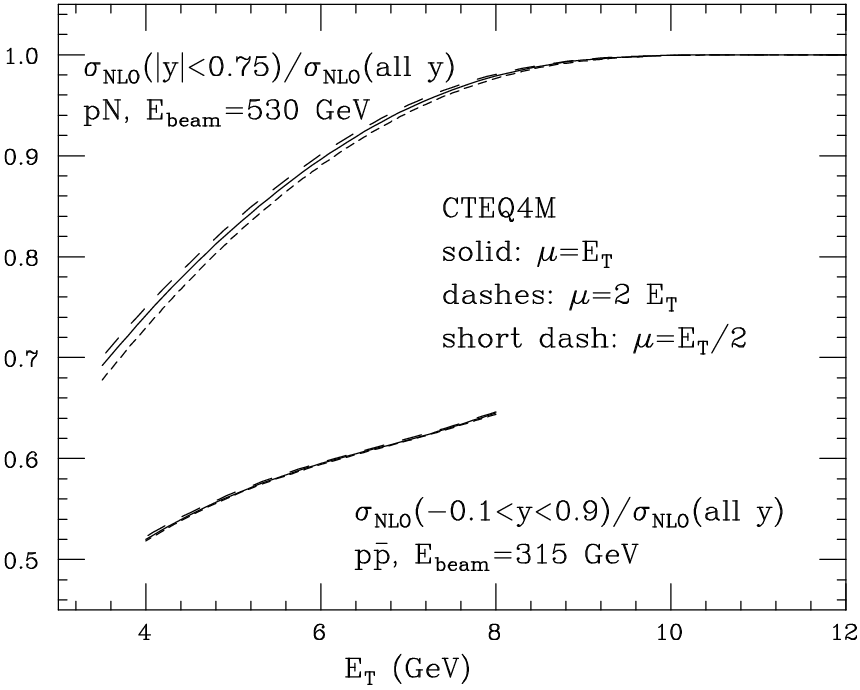,width=0.6\textwidth,clip=}
\vspace*{\mysep}      
\ccaption{}{\label{fig:yaccept} 
Effect of the rapidity cuts, as a function of \et. Upper set of curves:
E706. Lower set of curves: UA6. The three curves in each set are
obtained with $\mu=\et/2$, \et\ and $2\et$.}
\end{center}                                  
\end{figure}

As anticipated above, the comparison of our results with actual data 
requires the inclusion of realistic detector acceptance cuts. 
The resummation formalism discussed so far allows the evaluation of the
transverse-energy distributions integrated over the full range of rapidity for
the observed photon. This approximation is technically correct, provided the
measurement is performed within a finite range in rapidity including the value
of $y=0$ in the collision centre-of-mass frame. This is because in the
large-\xt\ limit all production is concentrated at $y=0$. 
To include the effect of experimental rapidity cuts, which usually do
include the $y=0$ point, we therefore apply the following acceptance
correction to our resummed cross sections:
\be
\sigma^{\rm (res)}(y\in Y) \equiv \sigma^{\rm (res)}({\rm all}~ y) \;
\frac{\sigma^{\rm (NLO)}(y\in Y)}{\sigma^{\rm (NLO)}({\rm all}~ y)} \;.
\ee
The experimental configuration for most experiments of practical interest is wide
enough that, for the relevant values of $\et$, the rapidity acceptance
is very large. We show two examples in Fig.~\ref{fig:yaccept}. 

One set of curves gives the ratio $\sigma(\vert y \vert
<0.75)/\sigma({\rm all}~ y)$, evaluated at NLO for the E706
experimental configuration.  The three curves correspond to different
choices of scale $\mu$, and show very small dependence on $\mu$. We
also checked that the dependence on the PDF set used is at the level
of 1-2\%.  The acceptance loss is of the order of 25\% for \et\ values
around 4~GeV, and becomes totally negligible for $\et\gtrsim 8$~GeV.
The other set gives the ratio $\sigma(-0.1< y<0.9)/\sigma({\rm
  all}~ y)$, evaluated at NLO for the UA6 experimental configuration.
The acceptance loss is here more significant, due to the tight cut at
negative rapidity.

\begin{figure}
\begin{center}
\epsfig{file= 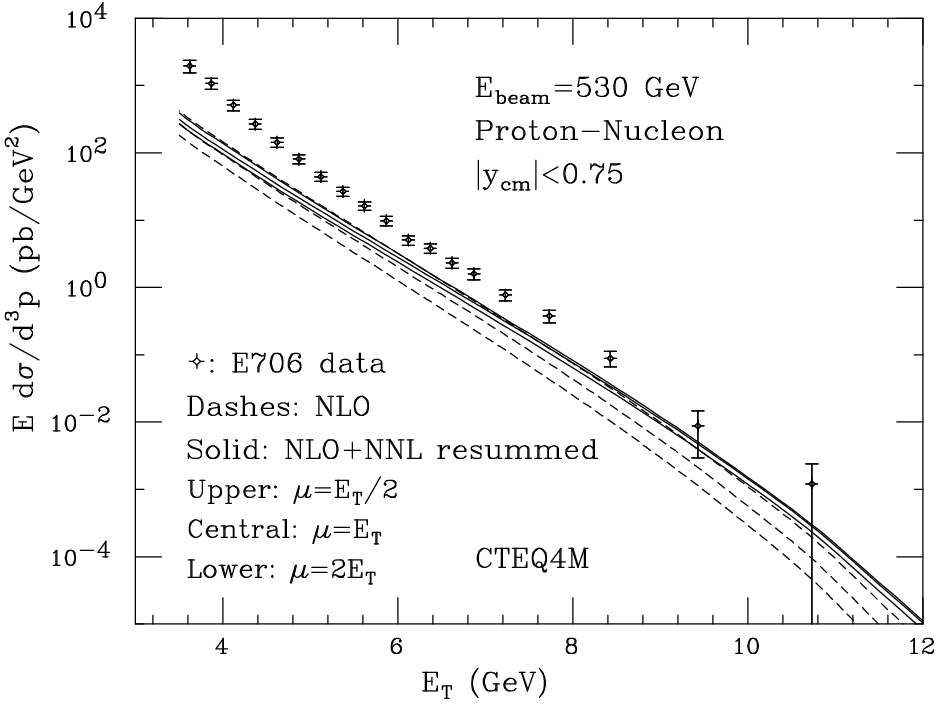,width=0.7\textwidth,clip=}
\vspace*{\mysep}      
\ccaption{}{\label{fig:e706comp} 
E706 data compared to the resummed theoretical predictions. The theory
was rescaled by a factor of 1.09 to account for nuclear corrections in
the conversion from the $p{\rm Be}$ to the $pN$ rate~\cite{Huston98}. 
For comparison, the figure includes as well the fixed-order NL
result. We used PDF set CTEQ4M, and GRV photon fragmentation
functions.}
\end{center}                                  
\end{figure}

We present our prediction for the E706 data in
Fig.~\ref{fig:e706comp}\footnote{The theoretical prediction for $pN$
  has been rescaled by a 1.09 factor, to account for nuclear
  corrections to the $p{\rm Be}$ process~\cite{Huston98}.}.  Notice
the significant reduction in scale dependence obtained when going from
the fixed-order NL calculation to the resummed result. This scale
reduction is particularly evident at high \et, where the resummation
effects are more important. Notice that while at low \et\ the band
with the resummed prediction is all contained within the NLO
uncertainty band, at high \et\ the NLL result becomes larger relative
to NLO for all the displayed scale choices. The plot shows a
reasonable agreement between data and theory at large \et, indicating
that no additional significant contribution is required in this region.
The large disagreement between data and theory already present at
NLO~\cite{Huston98} is still present, as no net increase is obtained
from the resummation contributions. Their only effect is to reduce the
scale dependence.

\begin{figure}
\begin{center}
\epsfig{file= 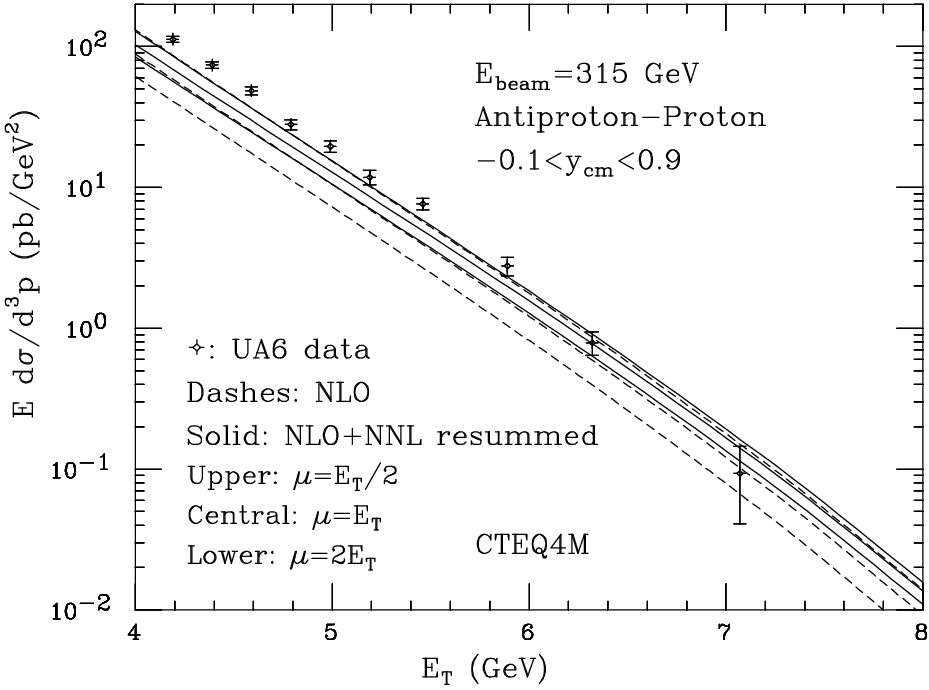,width=0.7\textwidth,clip=}
\vspace*{\mysep}      
\ccaption{}{\label{fig:ua6comp} 
UA6 ${\bar p} p$ data compared to the resummed theoretical predictions. 
For comparison, the figure includes as well the fixed-order NL
result. We used PDF set CTEQ4M, and GRV photon fragmentation
functions.}
\end{center}                                  
\end{figure}

A similar picture emerges from the comparison of the theory with the
${\bar p} p$ UA6 data. This is shown
in Fig.~\ref{fig:ua6comp}. The disagreement between data and
theory at low \et\ is however much less dramatic here than in the case
of E706. The extent to which these low-\et\ discrepancies can be
removed by the inclusion of non-perturbative effects such as an
intrinsic $k_T$ remains to be understood, as the global consistency of
the different data sets is not very
compelling~\cite{Aurenche98,Huston99,Martin98}.
                           
\section{Discussion and conclusions}
\label{conclusions}
We presented in this paper a numerical study of the impact of
resummation corrections, at the next-to-leading logarithmic level, on
the transverse energy distribution of direct photons produced in
hadronic collisions. We dealt with the resummation of the $\xt\to1$
Sudakov logarithms studied theoretically in
Refs.~\cite{CMN98,Laenen98}. As a result, this work is mostly of
relevance for typical fixed-target photon production. The current
prompt-photon data from the high-energy hadronic colliders cover in
fact the region $\xt\lesssim 0.1$, where Sudakov effects are negligible.

We showed that the inclusion of higher-order Sudakov corrections
improves significantly the factorization and renormalization scale
dependence, relative to what observed in the fixed-order NL
calculations. Even when the scales are varied independently, the
uncertainty from scale variations is significantly reduced. As a
result of the reduced scale dependence, the overall size of the
resummation contributions depends significantly on the chosen scale.
In general, however, the resummed cross sections for different scale
choices have values contained within the NLO uncertainty band for \xt\ 
values up to 0.5, and exceed the upper side of the NLO band by large
factors when \xt\ approaches 1. Still, our resummation corrections
turn out to be much smaller, at least in the range of existing data,
than the effects induced at large \xt\ by some implementations of
intrinsic $k_T$ effects, as discussed in
Refs.~\cite{Huston98,Huston99,Martin98}. In these papers the effect of
intrinsic-$k_T$ corrections was evaluated to be as large as factors of
3 and more, for the whole \et\ range.
We believe that this is related to the absence in the
intrinsic-$k_T$ models of the appropriate Sudakov suppression due to
the presence of the hadronic system recoiling against the photon
(represented at LL in our formalism by the negative terms in the
exponents of Eqs.~(\ref{dlresqg''}) and~(\ref{dlresqq''})).

In our work we did not include resummation corrections to the
fragmentation processes. We proved that, in $pN$ collisions, the
large-$N$ behaviour of the corrections to the $qq^{(\prime)}\to
qq^{(\prime)} \gamma$
processes is formally similar to that of the corrections to the
leading one, $qg\to q\gamma$. These corrections are therefore not
suppressed when $N$ increases towards larger values.
Whether they can be neglected or
not, is therefore a pure matter of numerics. We showed that their
contribution is not dominant in the \et\ regions of experimental
interest, and limited ourselves to including them at the fixed
next-to-leading order. As discussed in Sec.~\ref{fragcomp}, we have
no reasons to believe that the resummation corrections are any larger
for the fragmentation processes than for the $qg$ channel.  A more
quantitative study of these statements, and a complete
phenomenological assessment of the comparison between theory and the
current sets of data in view of the results presented in this paper,
will be the subject of future work.

\appendix
\section{Appendix: Formulae for the resummed cross section}
\label{appa}
In this Appendix we recall (see Ref.~\cite{CMN98}) the explicit
expressions of the various factors that contribute to the resummed
cross sections in Eqs.~(\ref{gammaresqq}) and (\ref{gammaresqg}).  We
use the customary notation for the colour factors in $SU(N_c)$ QCD,
namely, $C_F= (N_c^2-1)/(2N_c), C_A= N_c$ and $T_R=1/2$.

The $N$-moments of the LO partonic cross sections in 
Eqs.~(\ref{siqqgamma}, \ref{siqggamma}) are
\beq
\label{qgammaN}
{\hat \sigma}_{q{\bar q}\to g\gamma, \;N}^{(0)} = \pi \,e_q^2 
\,\frac{C_F}{N_c} \;\frac{\Gamma(1/2) \;\Gamma(N+1)}{\Gamma(N+5/2)}
\;(2+N) \;\;,
\eeq
\beq
\label{ggammaN}
{\hat \sigma}_{qg\to q\gamma, \;N}^{(0)} =
{\hat \sigma}_{{\bar q}g\to {\bar q}\gamma, \;N}^{(0)} = \pi \,e_q^2 
\,\frac{1}{8N_c} \;\frac{\Gamma(1/2) \;\Gamma(N+1)}{\Gamma(N+5/2)}
\;(7+5N) \;\;,
\eeq
where $\Gamma(z)$ is the Euler $\Gamma$-function.

The formulae for the NLL functions $g^{(2)}$ in the exponent of the 
radiative factors in Eq.~(\ref{deltanll}) are the following
\beeq
\label{g2qq}
g_{q{\bar q}}^{(2)}\!\left(\lambda,\frac{Q^2}{\mu^2};\frac{Q^2}{\mu_F^2}\right)
&=& (2C_F - C_A) \;h^{(2)}(\lambda) + 2 \,C_A \;h^{(2)}(\lambda/2)  \\
&+& \frac{2C_F - C_A}{2\pi b_0} \ln 2 \ln(1-2\lambda) 
+ \frac{ C_A \GE -  \pi b_0}{\pi b_0} \ln(1-\lambda) 
- \frac{2C_F}{\pi b_0} \;\lambda \ln \frac{Q^2}{\mu_F^2}
\nonumber \\
&+& \left\{ \frac{C_F}{\pi b_0} \Bigl[ 2\lambda + \ln(1-2\lambda) \Bigr]  
+ \frac{C_A}{2\pi b_0} \Bigl[ 2 \ln(1-\lambda) - \ln(1-2\lambda) \Bigr]
\right\} \ln \frac{Q^2}{\mu^2} \;, \nonumber
\eeeq
\beeq
\label{g2qg}
g_{qg}^{(2)}\!\left(\lambda,\frac{Q^2}{\mu^2};\frac{Q^2}{\mu_F^2}\right)
&=& C_A \;h^{(2)}(\lambda) + 2 \,C_F \;h^{(2)}(\lambda/2) \\
&+& \frac{C_A}{2\pi b_0} \ln 2 \ln(1-2\lambda) 
+ \frac{4 C_F \,\GE - 3 C_F}{4\pi b_0} \ln(1-\lambda) 
- \frac{C_F + C_A}{\pi b_0} \;\lambda \ln \frac{Q^2}{\mu_F^2}
\nonumber \\
&+& \left\{ \frac{C_F + C_A}{2\pi b_0} \Bigl[ 2\lambda + \ln(1-2\lambda) \Bigr]  
+ \frac{C_F}{2\pi b_0} \Bigl[ 2 \ln(1-\lambda) - \ln(1-2\lambda) \Bigr]
\right\} \ln \frac{Q^2}{\mu^2} \;, \nonumber
\eeeq
where $\GE = 0.5772\ldots$ is the Euler number and 
$b_0, b_1$ are the first two coefficients of the QCD $\beta$-function
\beq
\label{betass}
b_0 = \frac{11 C_A - 4 T_R N_f}{12\pi}\;,\;\;\;\;\; b_1 =
\frac{17 C_A^2 - 10 C_A T_R N_f -6 C_F T_R N_f}{24\pi^2}\;.
\eeq
The auxiliary function $h^{(2)}$ that appears in Eqs.~(\ref{g2qq}, \ref{g2qg}) 
has the following expression
\beq
\label{htnll}
h^{(2)}(\lambda) =
\frac{b_1}{2\pi b_0^3}\Bigl[ 2\lambda + \ln(1-2\lambda)
+\half\ln^2(1-2\lambda) \Bigr] 
- \frac{\GE}{\pi b_0} \ln(1-2\lambda) 
- \frac{K}{4\pi^2 b_0^2}\Bigl[ 2\lambda + \ln(1-2\lambda)\Bigr] 
\; , 
\eeq
where the coefficient $K$ is given by
\beq
\label{kcoef}
K = C_A \left( \frac{67}{18} - \frac{\pi^2}{6} \right) 
- \frac{10}{9} T_R N_f \;.
\eeq

The first-order coefficients $C_{q{\bar q}\to \gamma}^{(1)}$
and $C_{qg\to \gamma}^{(1)}$ of the $N$-independent functions in
Eq.~(\ref{cgamma}) are
\beeq
C_{q{\bar q}\to \gamma}^{(1)}(Q^2/\mu^2;Q^2/\mu_F^2) &=& 
\GE^2 \Bigl( 2 C_F - \frac{1}{2} C_A \Bigr)
+ \GE \Bigl[ \pi b_0 - (2 C_F - C_A) \ln 2 \Bigl] 
- \, \frac{1}{2} (2 C_F - C_A) \ln 2 \nonumber \\
\label{cqq1coef}
&+& \frac{1}{2} \,K - K_q 
+ \frac{\pi^2}{3} \Bigl( 2 C_F - \frac{1}{2} C_A \Bigr)
+ \frac{5}{4} (2 C_F - C_A) \ln^2 2 \\
&-& \Bigl( 2 \GE C_F - \frac{3}{2} C_F \Bigr) \ln \frac{Q^2}{\mu_F^2}
- \pi b_0 \ln \frac{Q^2}{\mu^2} \;\;, \nonumber
\eeeq
\beeq
C_{qg\to \gamma}^{(1)}(Q^2/\mu^2;Q^2/\mu_F^2) &=&
\GE^2 \Bigl( \frac{1}{2} C_F + C_A \Bigr)
+ \GE \Bigl[ \frac{3}{4} C_F -  C_A \ln 2 \Bigl]
- \, \frac{1}{10} ( C_F - 2 C_A) \ln 2 \nonumber \\
\label{cqg1coef}
&-& \frac{1}{2} \,K_q 
+ \frac{\pi^2}{60} \Bigl( 2 C_F + 19 C_A \Bigr)
+ \frac{1}{2} C_F  \ln^2 2 \\
&-& \Bigl( \GE (C_F + C_A) - \frac{3}{4} C_F - \pi b_0 \Bigr) 
\ln \frac{Q^2}{\mu_F^2}
- \pi b_0 \ln \frac{Q^2}{\mu^2} \;\;, \nonumber
\eeeq
where 
\beq
\label{kcoeffq}
K_q = \left( \frac{7}{2} - \frac{\pi^2}{6} \right) \,C_F \;\;,
\eeq
and the coefficient $K$ is given in Eq.~(\ref{kcoef}).

Note that the LL functions $g^{(1)}$ are given in Eq.~(\ref{g1fun}). Thus,
the formulae presented in this Appendix complete all the ingredients
that are necessary to evaluate the resummed cross sections with NLL accuracy.

\end{document}